\documentclass[useAMS,usenatbib]{mn2e}
\usepackage{graphicx} 
\usepackage{times}

\def\kms{km ${\rm s}^{-1}$}

\def\ch2{$\chi^2$}
\def\dg{$^{\circ}$}
\def\kms {\hbox{${\rm km\ s}^{-1}$}}

\def\ccm {$\hbox{{\rm cm}}^{-3}$}    %cm-3
\def\scm  {$\hbox{{\rm cm}}^{-2}$}    %cm-2
\def \HI {H{\sc \,i}}

\defcitealias{kc02}{KC03}

\def\lapp{\ifmmode\stackrel{<}{_{\sim}}\else$\stackrel{<}{_{\sim}}$\fi}
\def\gapp{\ifmmode\stackrel{>}{_{\sim}}\else$\stackrel{>}{_{\sim}}$\fi}
\def\bsp_small{\vspace{0.5cm}\small\noindent This paper has been typeset
from a \TeX/\LaTeX\ file prepared by the author.\normalsize}

\title[\HI\ 21-cm Absorption in DLAs]{Spin temperatures and covering
factors for \HI\ 21-cm absorption in damped Lyman-{\boldmath $\alpha$}
systems}
\author[S. J. Curran et al.]{S. J. Curran$^{1}$\thanks{E-mail:
sjc@phys.unsw.edu.au},  M. T. Murphy$^{2}$, Y. M. Pihlstr\"{o}m$^{3}$, J. K. Webb$^{1}$, 
C. R. Purcell$^{1}$\\
$^{1}$School of Physics, University of New South Wales, Sydney
NSW 2052, Australia\\ $^2$Institute of Astronomy, Madingley Road,
Cambridge CB3 0HA, UK\\ $^{3}$National Radio Astronomy Observatory, P.O. Box 0, Socorro, NM87801, USA}

\begin{document}

\date{Accepted ---. Received ---; in original form ---}

\pagerange{\pageref{firstpage}--\pageref{lastpage}} \pubyear{2004}

\maketitle

\label{firstpage}

\begin{abstract}
We investigate the practice of assigning high spin temperatures to
damped Lyman-$\alpha$ absorption systems (DLAs) not detected in
\HI~21-cm absorption. In particular, \citet{kc02} have attributed the
mix of 21-cm detections and non-detections in low redshift ($z_{\rm
abs}\leq2.04$) DLAs to a mix of spin temperatures, while the
non-detections at high redshift were attributed to high spin
temperatures. Below $z_{\rm abs}=0.9$, where some of the DLA host
galaxy morphologies are known, we find that 21-cm absorption is
normally detected towards large radio sources when the absorber is
known to be associated with a large intermediate (spiral)
galaxy. Furthermore, at these redshifts, only one of the six 21-cm
non-detections has an optical identification and these DLAs tend to
lie along the sight-lines to the largest background radio continuum
sources. For these and many of the high redshift DLAs occulting large
radio continua, we therefore expect covering factors of less than the
assumed/estimated value of unity. This would have the effect of
introducing a range of spin temperatures considerably narrower than
the current range of $\Delta T_{\rm s}\gapp9000$~K, while still supporting
the hypothesis that the high redshift DLA sample comprises a larger
proportion of compact galaxies than the low redshift sample.
\end{abstract}


\section{Introduction}\label{sec:intro}

Despite their relative paucity, the highest column density absorption systems,
damped Lyman-$\alpha$ systems (where $N_{\rm HI}\ge2\times10^{20}$
\scm), are an important component of the high redshift ($z \sim 2-6$)
Universe since they account for most of the neutral gas available for
star formation \citep[e.g.][]{lwt+91}. The \HI ~21-cm hyperfine
transition can provide an alternative and complementary view of
DLAs. Presuming that the 21-cm and Lyman-$\alpha$ absorption arise in the
same cloud complexes \citep{dl90}\footnote{Where the 21-cm and Lyman-$\alpha$
absorption trace the cold and total neutral hydrogen, respectively.},
the column density [\scm] of the absorbing gas in a homogeneous cloud
is related to the velocity integrated optical depth,
where $\tau\equiv-\ln\left(1-\frac{\sigma}{f\,S}\right)$, of the 21-cm line via
\citep{wb75}:
\begin{equation}
%N_{\rm HI}=-1.823\times10^{18}.T_{\rm spin}\int\!\ln\left(1-\frac {\sigma}{f.S}\right)\,dv\,,
N_{\rm HI}=1.823\times10^{18}\,T_{\rm spin}\int\!\tau\,dv\,,
\label{enew}
\end{equation}
where $T_{\rm s}$ [K] is the spin temperature, $\sigma$ is the depth
 of the line (or r.m.s. noise in the case of a non-detection) and $S$
 and $f$ the flux density and covering factor of the background
 continuum source, respectively.

As summarised by \citet{wol80}, 21-cm
 absorption line studies can face some difficulties compared with
 optical work: (1) in blind surveys the bandwidth of radio
 receiver/backend combinations only allows small redshift intervals to
 be searched for \HI ~absorption; (2) terrestrial interference can be
 severe below 1.4\,GHz; (3) if the spin temperature of the absorbing
 gas is high or if the background source size exceeds that of the
 absorption cloud(s), then the apparent optical depth can be
 considerably overestimated. Despite these technical difficulties,
 21-cm studies of DLAs have provided information about the physical
 conditions of absorbers, including the kinematics, temperature and
 velocity distribution of the gas.

One unresolved question concerns the typical size and structure of DLAs,
with models ranging from large, rapidly rotating proto-disks
(e.g. \citealt{pw97}) to small, merging sub-galactic systems
(e.g. \citealt*{hsr98}). Direct imaging by various groups
\citep[e.g.][]{lbbd97,rnt+03,cl03} of the low-$z$ DLA host-galaxies reveals
a mix of spirals, dwarf and low surface brightness (LSB) galaxies. This
`mixed morphology' picture of DLA host-galaxies appears to be reflected in
the nearby Universe in a recent 21-cm $z=0$ emission study \citep*{rws03}.

\begin{figure}
\vspace{8.8cm}
\includegraphics{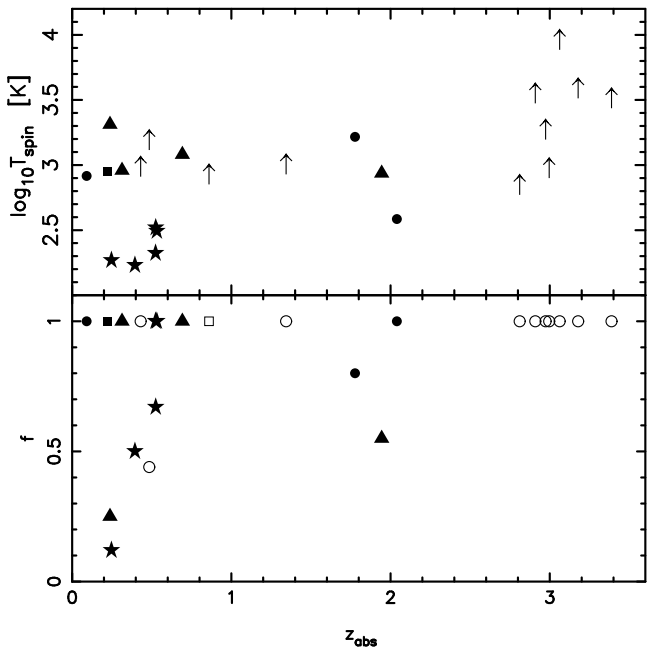}
\caption{Top: Spin temperature against absorption redshift as given in
\protect\citetalias{kc02}. The solid symbols represent the 21-cm
detections and the shapes represent the type of galaxy with which the
DLA is associated: circle--unknown type, star--spiral, square--dwarf,
triangle--LSB (obtained from \protect\citetalias{kc02} and references
therein). This key is also applied to Figs. \ref{m2}
%, \ref{m3},\ref{mike-sizes} and 
 to \ref{Toverf}. The arrows show the spin
temperature lower limits and all of these bar one (0454+039 at $z_{\rm
abs}=0.8596$) have unknown host identifications. Bottom: The covering
factors applied to derive the spin temperatures. The shapes are coded
as above with the solid and open symbols representing the covering
factors applied for the 21-cm detections and non-detections,
respectively.}
\label{kanekar}
\end{figure}

From a survey of 10 sources, in conjunction with the available literature,
\citet[][ hereafter \citetalias{kc02}]{kc02} find both detections and
non-detections of 21-cm absorption in DLAs at $z<2$, whereas the high-$z$
results are almost exclusively non-detections (Fig. \ref{kanekar}, top).
Additionally, many of the $z<1$ DLAs have host-galaxy identifications from
optical imaging, given in Table \ref{t2}. By combining the 21-cm optical
depth with the total \HI\ column density from optical spectroscopy
covering the damped Lyman-$\alpha$ line, \citetalias{kc02} find low spin
temperatures ($T_{\rm s}\approx 200$ K), which are typical of the Milky Way
and local spirals \citep{dl90}, in all cases where the absorber was
identified as a spiral (Fig. \ref{kanekar}). Furthermore, all DLAs at $z <
1$ with temperatures $T_{\rm s} \geq 1000$~K were found to be associated
with dwarf or LSBs and \citetalias{kc02} interpreted the non-detections at
high-$z$ ($>2$) as DLAs with high $T_{\rm s}$. They therefore advocate a
picture where the DLA host-galaxy population is dominated by the warmer
dwarfs and LSBs at high-$z$ and evolves to include a higher proportion of
spirals at low-$z$, thus satisfying one expectation of hierarchical galaxy
formation scenarios. However, in many of these studies, particularly at
high redshift, the covering factor, $f$, of the background continuum is
often assumed to be unity (Fig. \ref{kanekar}, bottom -- see also Table
\ref{t2}.). Since $f$ plays an equal r\^{o}le as $T_{\rm s}$ in relating the
column density to the observed optical depth profile, it is important to
consider the robustness of this assumption, which is the aim of this
article.

%\section{Search for 21-cm absorption in three southern DLAs}\label{sec:obs}
\section{Factors affecting the detection of 21-cm absorption}
\label{sec:biases}

\subsection{Previous 21-cm absorption searches}
\label{prev}
In Table \ref{t2} we summarise the previously published 21-cm searches to
which we add the 3 non-detections detailed in Appendix A. The optical
depths are quoted for various resolutions between 0.6 and 17 \kms.
\begin{table*}
\centering
\begin{minipage}{170mm}
\caption{The DLAs and sub-DLAs which have been searched for 21-cm
absorption. In the top panel we list the detections and in the bottom
panel the non-detections. $\tau_{\rm peak}$ is the peak optical depth
of each individually resolved component, the number of which are
listed under $n_{\rm comp}$ (B designates a blend) and $\Delta v$ is the full
width half maximum (FWHM) of the line [\kms]. For the non-detections the
$3\sigma$ upper limits of $\tau_{\rm peak}$ at a velocity resolution
of 3 \kms (see main text) are quoted. $f$ and $T_{\rm s}$ are the
covering factor used and the inferred spin temperature [K] from the
literature, respectively. For the former $^{\dagger}$denotes that $f$
is estimated from high resolution radio images and $^*$denotes that
$f$ is assumed. $N_{\rm HI}$ and $z_{\rm abs}$ are the total neutral
hydrogen column density [\scm] (see
http://www.phys.unsw.edu.au/$\sim$sjc/dla) and the redshift of the
DLA, respectively, with the optical identification (ID) given:
D--dwarf, L--LSB, S--spiral, U--unknown. $z_{\rm em}$ is the redshift
of the background QSO and $S$ is the flux density [Jy] at $\nu_{\rm
obs}$, obtained from either the reference given or as cited in
Curran et al. (2002b). The final column gives the 21-cm absorption
reference. \label{t2}}
\begin{tabular}{@{}l c c c c c c  c c cc l c@{}} 
\hline
QSO & $\tau_{\rm peak}$& $n_{\rm comp}$ &$\Delta v$  & $f$ & $T_{\rm spin}$ & $\log N_{\rm HI}$  & $z_{\rm abs}$ & ID & $z_{\rm em}$ &$S$ & Ref. \\
\hline
0235+164 & 0.7/0.5& 3B+1 & -- & 1$^{\dagger}$ & 100 & 21.7 & 0.52385  & S & 0.940 & $\approx1.7$ &  2\\
0248+430 & 0.2/0.2/0.1/0.1& 4 &-- & 1$^*$ & --& -- & 0.394 & U & 1.31 & 1.0 & 22\\
0458--020  & 0.3--3.1& 2B & 25 & 1--0.28$^{\dagger}$ & $<1000$ &21.7  & 2.03945 & U &2.286 & 3 & 7\\
0738+313 & 0.04& 1 & 8 & 1$^{\dagger}$ & 1100 &20.8 & 0.2212& D & 0.635 &1.9 & 14\\ 
...& 0.25 & 2 & 4 & 0.98$^{\dagger}$ & 800 & 21.2 & 0.0912 & U & ... & 2.2 & 19\\
0809+483$^{a}$ & 0.024 & 2B & 35 & -- & -- &20.2 & 0.4369 & S &0.871 & 19.0& 5\\
0827+243 & 0.007&1 & 50 & 0.67$^{\dagger}$ & 300 & 20.3 & 0.5247 & S & 0.939 & 0.9 & 20\\
0952+179 &0.013& 1 &8 & 0.25$^{\dagger}$ & 2000 & 21.3 & 0.2378 & L & 1.472 & 1.4  & 20\\
1127--145 & 0.06,0.09& 4B+1 &42 & 1,1$^{\dagger}$ & 1000,910& 21.7 &0.3127 & L & 1.187 & 5.3,6.2 & 14,18\\
1157+014 & 0.05--0.2&1 &42 & 0.25--1$^{\dagger}$ & -- & 21.8 & 1.94362 &L & 1.986 &  0.80 &4\\
1229--021 & 0.05 &2B & 5&0.43$^{\dagger}$ & -- & 20.8 &0.39498 & S & 1.045 & 2.2 & 6,13\\
1243--072 & 0.07 &1&  14 & 1$^*$& -- & -- & 0.4367 & S & 1.286 & 0.48 & 22\\
1328+307$^{b}$ & 0.11 & 1&8& 0.2$^{\dagger}$ & 100 & 21.3 & 0.692154 & S & 0.849 & 19.0&  1\\
1331+170 & 0.02 & 1& 22 & 1$^*$ & 980 & 21.2 & 1.77642& U & 2.084 &  0.61 &3 \\
1413+135$^{c}$ & 0.3 &1& 18 & $\leq0.1$$^{\dagger}$ & -- &-- & 0.24671 & S &  0.24671 &1.25 & 8 \\
1629+120 & 0.039 &2B & 40 & 1$^{\dagger}$ & 20--310 & 20.7 & 0.5318 & S & 1.795 & 2.35 & 23\\
2351+456 & 0.32 &1& 53 & -- & -- & -- & 0.779452 & U & 1.9864 & 2.0 & 24\\
\hline
0118--272 & $<0.009$ & --& -- & 0.5$^{\dagger}$ & $>850$ & 20.3 & 0.5579 & U & 0.559 & 1.1 &20\\
0201+113 &0.09,0.04,$<0.09$$^{d}$ &1& 9,23,--&1$^{\dagger}$ & $1100, \ge5000$ & 21.3 &3.386  & U & 3.610 & 0.35 & 11,12\\
0215+015 & $<0.006$ & --&--& 1$^{\dagger}$ & $>1000$ & 19.9 &1.3439 & U & 1.715 & 0.92 & 23\\
0335--122 & $<0.008$ & --& -- & $\approx1^*$ & $>3000$ &20.8 & 3.178 & U & 3.442 & 0.68&23\\
0336--017 & $<0.007$& --& -- & 1$^*$ & $>9000$ &21.2 & 3.0619 & U & 3.197 &  0.94& 23\\ 
0432--440& --& --& --& --& --& 20.8 & 2.297& U & 2.649 & -- &25\\ 
0438--436 & $<0.1$ & --&-- &-- &-- & 20.8 & 2.347&U & 2.852 &  -- &25\\ 
0439--433 & $<0.012$ & --& --& 1$^*$ & $>700$ & 20.0  & 0.10097& U & 0.593 & 0.33 &21\\
0454+039 & $<0.02$& --& -- & 0.28$^{\dagger}$ & -- & 20.7 &0.8596  & D & 1.345 & 0.44 &6 \\
0528--250 &$<0.3$ &--& --& 1$^*$ & $>700$ &21.3  & 2.8112& U &2.779 & 0.14 &10\\
0537--286 & $<0.007$ & --& -- & 1$^{\dagger}$ & $>1900$ & 20.3 & 2.974& U& 3.104 & 1.1& 23\\
0906+430$^{e}$ & $<0.01$ & --& -- & -- & -- & -- & 0.63 & U & 0.670 & 6.14 & 17\\
0957+561A & $<0.01^f$ & --& -- & --& -- &20.3  & 1.391& U & 1.413 & 0.59&23\\
%C057 & 18.9& $<0.004$ & --& 20.1 &3.42  & 2.2 & &22\\
1225+317 & $<0.09$ & --&--& 0.11$^{\dagger}$ & -- &--  &1.821$^g$ & U &2.219 & 0.35 & 6\\
1228--113& --& --& --&-- & --& 20.6 & 2.193& U & 3.528 & -- & 25\\
1354--107 & $<0.07$& --&-- &1$^*$ & $>1000$ & 20.8  & 2.996& U & 3.006 & 0.12 &23\\
1354+258 &$<0.015^f$ & --& --& -- & -- & 21.5& 1.4205 & U & 2.006 & 0.30 & 23\\
1451--375 & $<0.007$ & --&--& 1$^*$ & $>1400$ & 20.1  & 0.2761&U & 0.314 & 3.8$^h$  & 18\\
2128--123 & $<0.003$ & --& -- &$\approx1^{\dagger}$ & $>1000$ &  19.4 &0.4298  & U & 0.501 & 1.9 & 23\\
2223--052$^{i}$ &$<0.03$ & --&-- & 0.44$^{j}$ & $>1600$ & 20.9 &0.4842  &U & 1.4040 & 8.0$^h$   &18\\
2342+342 &$<0.03$  & --& --& 1$^*$ & $>4000$ & 21.3 & 2.9084 & U &3.053 & 0.31 &23\\\hline
\end{tabular}
{Notes: $^{a}$3C196, $^{b}$3C286, $^{c}$included by \citetalias{kc02},
but since no Lyman-$\alpha$ absorption has been detected in this
associated system we shall exclude it from our analysis, $^{d}$also,
\citet{bbw97} detected 21-cm absorption with the WSRT but not the VLA;
as with the current consensus (e.g. {Wolfe}, {Gawiser} \& {Prochaska}
2003) we shall consider this a non-detection, $^{e}$3C216 (although
not detected in the Lyman-$\alpha$ line nor Mg{\sc \,ii} 2796/2803 \AA
~doublet, the UV spectrum of \citet{wth+95} suggests a possible DLA),
$^{f}$these $3\sigma$ upper limits are not quoted in \citetalias{kc02}
since the low core radio fluxes reduce the optical depth sensitivity
as well as implying a low covering factor in both cases,
$^{g}$observed at $z=1.795$, $^{h}$flux density estimated from
neighbouring frequencies (see
http://www.phys.unsw.edu.au/$\sim$sjc/dla/), $^{i}$3C446,
$^{j}$although $f=0.1$ is quoted by \citetalias{kc02} we find this to
be 0.44 from \citet{ck00}. \\ References: (1) \citet{br73}, (2)
\citet{rbb+76}, (3) \citet{wd79}, (4) \citet{wbj81}, (5) \citet{bm83},
(6) \citet{bw83}, (7) \citet{wbt+85}, (8) \citet{cps92}, (9)
\citet{lwt95}, (10) \citet{cld+96}, (11) \citet{dob96}, (12)
\citet{kc97}, (13) \citet{bri98}, (14) \citet{lsb+98}, (15)
\citet{rt98}, (16) \citet{ck99}, (17) \citet{pvtc99}, (18)
\citet{ck00}, (19) \citealt{lbs00}, (20) \citet{kc01a}, (21)
\citet{kcsp01}, (22) \citet{lb01}, (23) \citetalias{kc02}, (24)
\citet{dgh+04}, (25) This paper (see Appendix A).}
\end{minipage}
\end{table*}
As discussed in detail by Curran et al. (2002a), the choice of resolution
affects the upper limit to the optical depth obtained and so we
normalise these to 3 \kms, a fairly typical resolution for the DLAs
detected in 21-cm absorption.  In the table we also give the covering
factors used to determine the spin temperature in each of the previous
searches. Where no high resolution radio images are available for both
the background source and absorber, it is common practice to assume
that the absorber completely covers the continuum source
(e.g. \citealt{lb01}), although in some cases the covering factor is
estimated as the ratio of the compact unresolved component's flux to
the total radio flux (e.g. \citealt{bw83}). A good example of this
method is given by \citet{lbs00}, where the constancy of the 21-cm 
absorption profile in the $z_{\rm abs}=0.09$ DLA over the partially resolved 
core of 0738+313 gives an estimate of $f\approx0.98$. As \citet{bw83}
mention, however, even though this method provides information on the
background source structure, it provides none on the size of the
absorbing region\footnote{The effect of background source size is
clearly demonstrated, for instance, in absorption studies of PSR
B1849+00 \citep{swd+03a}.}. The only way to unambiguously determine
the covering factor is by mapping both the source and the absorbing
gas at high angular resolution, which is challenging due to the very
high sensitivities required.

\subsection{Column density and background flux biases}
\label{20kms}

In order to check that the non-detection of 21-cm absorption in DLAs
is not the result of lower atomic hydrogen column densities, we plot
\begin{figure}
%	\vspace{8.3cm} \setlength{\unitlength}{1in}
\vspace{7.85cm}
\includegraphics{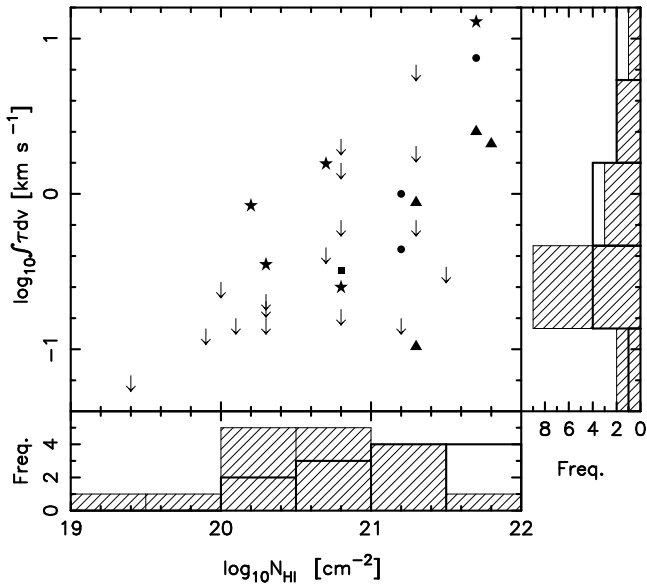}
\caption{The velocity integrated optical depth versus total neutral
hydrogen column density for the DLAs searched for 21-cm absorption. In
this and the following figure, the shapes/bold histogram represent the
detections and the arrows/hatched histogram the upper limits.}
\label{m2}
\end{figure}
the velocity integrated optical depth against column density in
Fig. \ref{m2}.  We use this rather than the peak optical depth as this
provides a better representation of the strength of the line. For a
single cloud in thermodynamic equilibrium, the spin temperature of the
gas can be estimated from the kinetic temperature according to $T_{\rm
spin}\approx T_{\rm kin}\lapp22\times{\rm FWHM}^2$
(e.g. \citealt{lb01}) and this has been used by \citetalias{kc02} in
order to derive spin temperatures and line-widths which are self
consistent. However, in the absence of detailed knowledge of the
covering factor, it is impossible to assign a specific spin
temperature to the system. We therefore assume a FWHM of 20 \kms (the
mean value of the detections) for the non-detections (Table
\ref{t2}). From the table see that the FWHM can range from 4 to 50
\kms, each of which would not cause too large a deviation from our
chosen value of 20 \kms ~on the log plots. This is confirmed by the
fact that using the peak rather than velocity integrated optical
depths gives similar qualitative results in Figs. \ref{m2} and
\ref{m3}.

Although the column densities for the non-detections are
generally lower, they appear to have been searched to correspondingly lower
optical depth limits. To check this, in Fig. \ref{m3}
\begin{figure}
%	\vspace{9.7cm} \setlength{\unitlength}{1in}
\vspace{7.85cm} \setlength{\unitlength}{1in}
%\special{psfile=modtau_vs_S.ps hoffset=-10 voffset=-60 hscale=40 vscale=40 angle=0}
\includegraphics{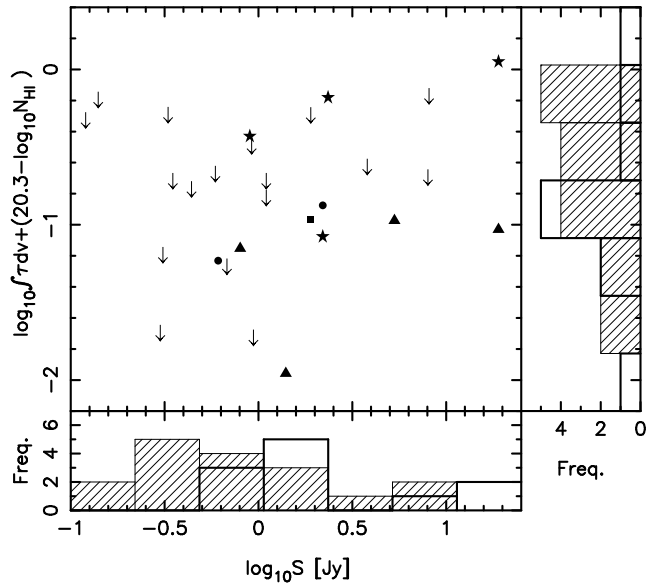}
\caption{The modified velocity integrated optical depth versus the radio flux density at
$\nu_{\rm obs}$ for the optically thin ($\tau\lapp0.3$) 21-cm searches.} 
\label{m3}
\end{figure}
we show the distribution of the velocity integrated optical depth
modified according to the column density of each DLA: In the optically
thin regime ($\sigma\ll f.S$) equation \ref{enew} reduces
to\footnote{As seen from Table \ref{t2}, this is a reasonable
assumption providing that $f$ is not small. For $f\lapp\frac{\sigma}{0.3S}$ the
assumption is not valid, although it supports the argument that $f\ll1$.}
\begin{equation}
N_{\rm HI}=1.823\times10^{18}\frac{T_{\rm spin}}{f}\int\!\frac {\sigma}{S}\,dv\,,
\label{e1}
\end{equation}
and from Table \ref{t2} we see that this is applicable
($\tau\lapp0.3$) to all but two systems, 0235+164 and 0458--020. With
this approximation we can introduce a linear correction to the
optical depth in order to account for the range of column densities
exhibited by the DLA sample.  Here we correct the optical depth by a
factor of $2\times10^{20}/N_{\rm HI}$, where $N_{\rm
HI}\geq2\times10^{20}$ \scm ~is nominally defined as the limiting
column density of a DLA. This has the effect of putting the data
points on an equal footing on the ordinate, i.e. for a given optical
depth limit we account for the fact that a non-detected system of low
column density has not been searched for as deeply as one of higher
column density. From Fig. \ref{m3} we see, not surprisingly, that
detections tend to occur towards higher flux densities, although the
samples are not mutually exclusive. For the modified peak optical
depths the overlap is more pronounced than in Fig. \ref{m3}, and for
both samples $-3\lapp\log_{10}\tau + [20.3-\log_{10}N_{\rm
HI}]\lapp-1.5$. Equation \ref{e1} therefore suggests similar values of
$f/T_{\rm spin}$ (over a 2 order of magnitude range) for the
detections and non-detections. We discuss the possible contributions
of $f$ and $T_{\rm spin}$ in the following section.

\subsection{Spin temperatures and covering factors}

From Section \ref{20kms} it appears that neither the column density
nor background flux (providing they are sufficiently high) are the
main factors in determining whether a DLA is detected in 21-cm
absorption.  As previously discussed, the popular consensus is
the wide range of estimated  spin temperatures
(possibly $20 ~{\rm to}~ > 9000$ K) in DLAs. In view that
the equally influential covering factor is often assumed to be unity,
we note the following:

\begin{enumerate}
\item In radio quasar hosts
	  the detection rate for 21-cm absorption is much higher in
	  the compact sources, where the coverage is high
	  \citep{pcv03,vpt+03}.
\item The presence of H$_2$ in 0528--020 \citep{lv85} suggests that
the gas is relatively cold with $T_{\rm kin}\approx 200$ K
(\citealt{sp98}), cf. $T_{\rm spin}>700$ K from a 21-cm non-detection
(\citealt{cld+96}).  This of course could be the result of different
sight-lines being probed as well as the possibility the absorption is
not due to a single cloud. This is known to be a complex situation
with $z_{\rm abs}\sim z_{\rm em}$ (Table \ref{t2}), suggesting an
infalling system rather than a simple intervener, although it is one
of the cases where a covering factor of unity is assumed. This is a
valid assumption, however, if the absorbing material has an extent of
$\gapp70$ pc.\footnote{From the radio continuum size of $0.01''$ (see
Table \ref{ylva}) at $z=2.8$ ($H_{0}=75$~km~s$^{-1}$~Mpc$^{-1}$,
$\Omega_{\rm matter}=0.27$).}

Note that a similar situation is found towards 0738+313, where $T_{\rm
kin}\lapp300$ K, cf. $T_{\rm spin}\approx800$ K in both absorbers
\citep{lbs00,kgc01}. As seen from the absorption profiles towards the
1.3 GHz VLBA image (figure 2 of \citealt{lbs00}), although overall $f$ is
large, the coverage towards the north appears to be considerably less
than towards the south-east. This suggests that the absorber extends
to $\lapp16$ pc and $\approx22$ pc in these
respective directions. A lower coverage in the northerly direction
could possibly reduce the spin temperature estimates to closer to the
kinetic temperature values, although the two-phase gas model suggested
by \citet{lbs00} offers an alternate explanation. In this model, the deep,
narrow absorption lines give the temperature of the cold gas, and the
much wider weaker line giving the warm gas temperature. These models
are discussed further at the end of this section.

\item
For the DLAs in which 21-cm absorption has been detected there appears
to be little correlation between spin temperature and
line-width. However, these are complex, multi-component systems
characterised by a single velocity integrated optical depth and
considering the possibility of both narrow/cold and wide/warm
components in the profile (above), the lack of correlation is not
surprising.  Such simplification does however demonstrate the possible
pitfalls in estimating line-widths for the non-detections (Section
\ref{20kms}).
\end{enumerate}

\begin{figure*}
	\vspace{13.4cm} \setlength{\unitlength}{1in}
\includegraphics{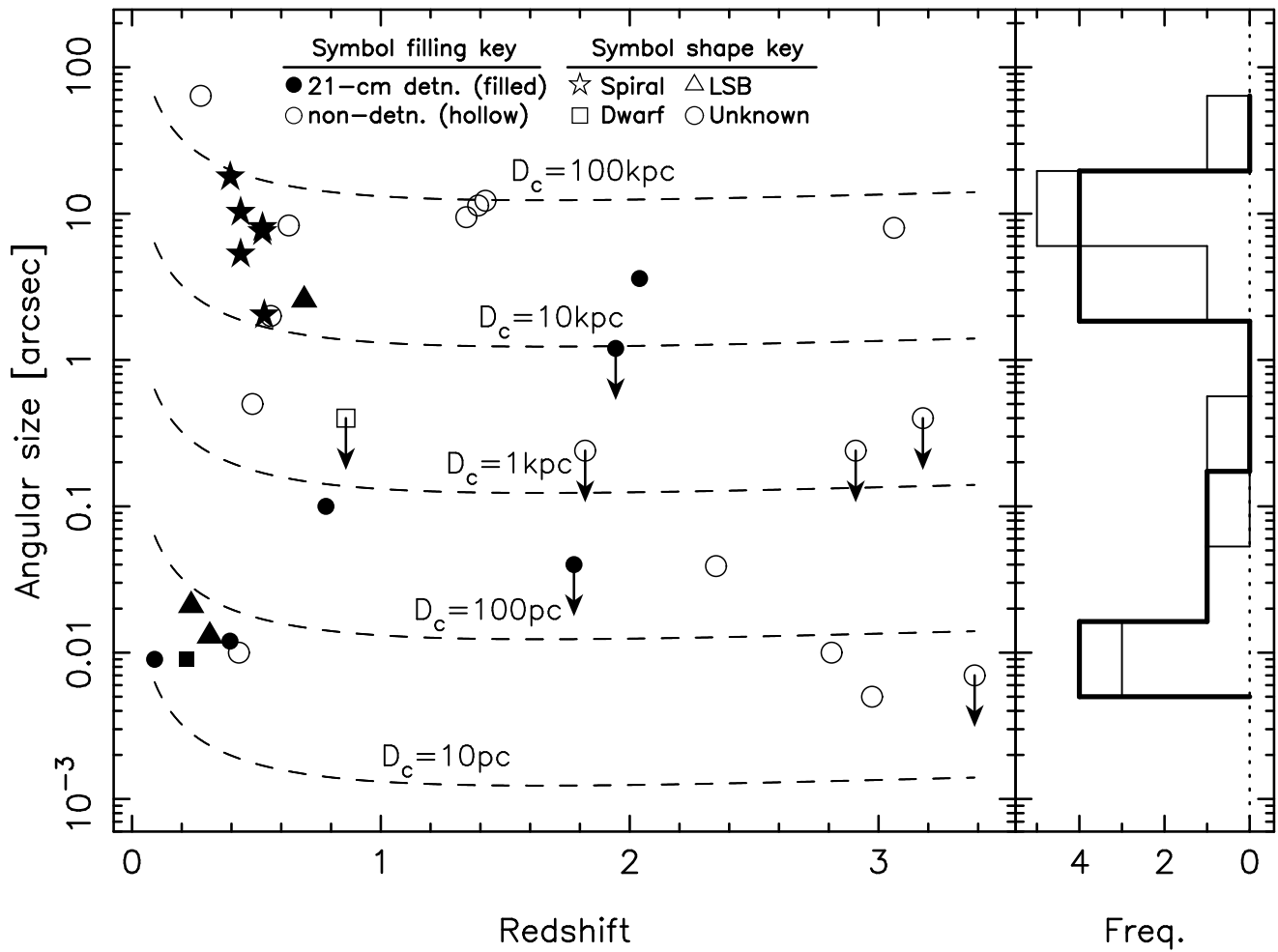}
\caption{The angular sizes of the background radio continuum
sources. Left: The dashed lines represent the angular extent of
different size absorption clouds (D$_{\rm c}$) as a function of redshift
($H_{0}=75$~km~s$^{-1}$~Mpc$^{-1}$, $\Omega_{\rm matter}=0.27$ and
$\Omega_{\Lambda}=0.73$). The source sizes are plotted at the absorption
redshift. The arrows represent continuum source upper limits, the
open symbols are the 21-cm non-detections and the solid symbols are the
detections.
% with the shapes representing the type of galaxy to which the
%DLA is associated: circle--unknown type, star--spiral, square--dwarf,
%triangle--LSB [as \protect\citetalias{kc02}]. 
Right: The frequency of
various source sizes, where the bold histogram represents the detections
(upper limits are excluded).}
\label{mike-sizes}
\end{figure*}

All 21-cm absorption detections occur in lower redshift ($z_{\rm
abs}\leq2.04$) DLAs (see Fig. \ref{mike-sizes}) and \citetalias{kc02} attribute this
to these having a mix of both low and high spin temperatures
(Fig. \ref{kanekar}, top), while the higher redshift DLAs generally have
high spin temperatures. This may imply that the more distant DLA sample
consists of a particular class of object. However, due to their large
distances, none of the high redshift 21-cm sample have hosts which have
been optically identified and indeed only one of the $z<2$ non-detections
has an optical identification (Fig. \ref{kanekar}, bottom). Therefore, we
have insufficient information with which to determine the DLA host
morphology and thus the likely covering factor.  From this and the above
points, we cannot rule out that perhaps the inferred (high) spin
temperatures are an artifact of an overestimate of the covering factor.

One way to address this issue is to estimate how the expected size of
the intervening absorber compares to that of the background continuum:
In Table \ref{ylva} we show the results from high resolution radio
images for the sources searched\footnote{Note that the continuum
source sizes estimated by \citetalias{kc02} from the spectral energy
distributions (SEDs) agree qualitatively with the results in the
table.  Specifically, the flat SEDs for 0215+015, 0335--122 \&
2342+342 suggest compact sources. For the remainder of their sources
they use the ratio of the unresolved component's flux (Section
\ref{prev}) to estimate the covering factor and, in order to avoid
erroneous spin temperature estimates, exclude from further analysis the
two sources (0957+561 \& 1354+258) where this is expected to be low.}
\begin{table*}
\centering
\begin{minipage}{157mm}
\caption{Radio source sizes (arc-seconds) and VLBI and VLA
morphologies of the QSOs illuminating DLAs searched for 21-cm absorption. \label{ylva}}
\begin{tabular}{@{}l  c r c  r c r l @{}} 
\hline
QSO & $\theta_{\rm QSO}$ & Ref & VLBI & Ref & VLA & Ref & Comments [Ref]\\
\hline
0235+164 & 7.5 & 16 & CJ & 25 & T & 16 & VLBI (5 GHz)/VLA (2 GHz) triple + extra strong component\\
0248+430 & 0.012 & 22 & T & 22 & U & 1 & VLBI (1.6 GHz) weak core + 2 strong lobes or a CJ structure, \\
&  & & & & & & outer J stronger than inner/VLA (5 GHz) $<0.8''$\\
0458--020 & 3.6 & 38 & CJ & 30 & CDom &38 &  VLBI (5 GHz)/VLA (5 GHz) core + weaker lobes \\
&  & & & & & & (2 component, 1.4 GHz VLA [11])\\
0738+313 & 0.009 & 29 & CJ/T & 29 & T & 16 & VLBI (5 GHz)/VLA (1.6 GHz) triple, FRI like (core strongest)\\
0809+483 & 5.27 & 3 & -- & -- & D & 3 & VLA (5 GHz) FRII like (strong lobes), as MERLIN 408 MHz [23]\\
0827+243 & 8.00 & 17 & CJ & 39 & D & 17 & VLBA (2 \& 8 GHz)/VLA (1.4 GHz) FRII like\\
0952+179 & 0.021 & 39 & T & 39 & --& -- & VLBA (2 \& 8 GHz) either CJ or C-J-CJ (one side stronger)\\
1127--145 & 0.013 & 24 & CJ & 24 & U & 1 & VLBI/VLA (1.4 \& 5 GHz)\\
1157+014 & $<1.2$ & 6 & -- & -- & U & 6 & VLA (5 GHz)\\
1229--021 & 17.76 & 4  & -- & -- & T & 4 & VLA standard triple, C + 2 weaker lobes\\
1243--072 & 10.17 & 5 & CJ & 39 & T & 5 & VLBA (2 \& 8 GHz)/VLA standard triple, core is dominant\\
1328+307 & 2.57 & 14 & CJ & 39 & CDom & 14 & VLBA (2 \& 8 GHz)/VLA (1.5 GHz) strong core + weaker component\\
1331+170 & $<0.04$ & 9,31 & CJ & 41 & U &9 & VLBA (2 \& 8 GHz)/VLA (5GHz)\\
1413+135$^*$ & 0.040 & 20 & CJ & 20 & U & 1 & VLBI CJ + 2 more extended components on each side, core stronger\\
&  & & & & & &VLA (5 GHz)\\
1629+120 & 2.025 & 15 & D & 32 & T & 15 & VLBI (2 GHz) EVN/VLA 3 strong components, possibly FRII\\
2351+456 & 0.03 & 22 & CJ & 22  & U&1 & VLBI (1.7 GHz) strong 20--30 mas core + one weaker 100 mas jet \\
\hline
0118--272 & 2.0 & 1 & T/mul & 34 & T/mul & 2 & VLBI (5 GHz) core + extended\\
&  & & & & & &VLA (5 GHz) at least 3 components, complex\\
0201+113 & $<0.007$ & 32 & C & 32  & CDom & 12 & Core + weak extended (1\% of total 21-cm flux within $2''$ [12])\\
0215+015 & 9.45 & 7 & C & 27 & T & 7 & VLBA (2 \& 8 GHz)/ VLA (5 GHz) FRI, core dominated\\
0335--122 & $<0.4$ & 11 & C & 41 & U & 11 & VLBI (2 GHz)/VLA (1.4 GHz)\\
0336--017 & 8.0 & 37 & CJ & 18& T  & 37 & VLBI (5 GHz) unresolved/VLA (5 GHz) FRI, core dominated\\
0432--440$^*$ & -- & -- & -- & --& -- & -- & --\\
0438--436 & 0.039 & 35 & D & 35 & U & 1 &  VLBI (5 GHz)/VLA (1 \& 5 GHz)\\
0439--433 & --& --& -- & -- & --& -- & --\\
0454+039 & $<0.4$ & 11 & CJ & 41 & U & 11 & VLBA (2 \& 8 GHz)/VLA (1 \& 5 GHz)\\
0528--250 & 0.010 & 40 & C & 40 & U & 1 & VLBI (5 GHz)/VLA (1 \& 5 GHz) \\
0537--286 & 0.005 & 40& CJ & 40& --& -- & VLBA (2 \& 8 GHz)\\
0906+430 & 8.3 & 14 &CJ & 22 & T & 14 & VLBI (1.6 GHz) core strongest/VLA (1.4 GHz) core strongest\\
0957+561 & 11.36 & 28 & CJ & 21 & T? & 28 & VLBI (5 GHz)/VLA (5 GHz) lensed, looks like CLL core strongest\\
1225+317 & $<0.24$ & 8 & C & 41& U & 8 & VLA (2 GHz) possibly CJ/VLA (5 GHz)\\
1228--113$^*$ & 0.01 & 41 & CJ & 41 & -- & -- & VLBA (2 \& 8 GHz)/VLA: no map, \\
&  & & & & & & may have flux on larger than VLBI scales\\
1354--107 & --& --& -- & -- & --& -- & --\\
1354+258 & 12.24 & 15 & --& -- & T & 15 & VLA (5 GHz) FRI like, core dominates\\
1451--375 & 63.79 & 33 & CJ & 39 & T & 33 & VLBA (5 GHz)/VLA (1.4 GHz) FRI like, core dominates\\
2128--123 & 0.010 & 26 & CJ & 26 & U & 1 & VLBA (2 \& 8 GHz)/VLA (1 \& 5 GHz)\\
2223--052 &0.5 & 13 & CJ & 40 & U & 4 & VLBA (2 \& 8 GHz)/VLA (1.4 GHz)  \\
2342+342 & $<0.24$ & 36 & -- & -- & U & 36 & VLA (8 GHz)\\
\hline
\end{tabular}
{Key VLBI morphology: CJ--core-jet morphology (most flux in core),
D--double (2 lobes, no core visible), C--core, T--triple (core plus
two lobes), mul = multiple components ($>3$). N.B. it is not always
possible to determine the core in the VLBI images.\\ Key VLA
morphology: U--unresolved, T--triple, usually the morphology is either
FRI like (most flux in core) or FRII like (most flux in lobes),
D--double, usually FRII like (core may be too faint), CDom--core
dominated, usually a very strong core plus some fainter emission
(some may look like very weak lobes). $^*$Note that these are included
in order to provide a comprehensive list but not included in the
analysis (Fig. \ref{mike-sizes}) -- 1413+135 is an associated system
and the limits for 0432--440 and 1228--113 are poor (Appendix A) \\
References: (1) \citet{ujpf81}, (2) \citet{per82}, (3) \citet{skn82},
(4) \citet{huo83}, (5) \citet{gh84}, (6) \citet{sfwc84}, (7)
\citet{au85}, (8) \citet{rpd87}, (9) \citet{bmsl88}, (10)
\citet{bwl+89}, (11) \citet{nh90}, (12) \citet{sbom90}, (13)
\citet{fpa92}, (14) \citet{vff+92}, (15) \citet{lbm93}, (16)
\citet{mbp93}, (17) \citet{pgh+93}, (18) \citet{gsb+94}, (19)
\citet{lgh94}, (20) \citet{pss+94}, (21) \citet{clc+95}, (22)
\citet{pwx+95}, (23) \citet{rsa+95}, (24) \citet{bpf+96}, (25)
\citet{cbr+96}, (26) \citet{fcf96}, (27) \citet{fc97}, (28)
\citet{hsmr97}, (29) \citet{sob+97}, (30) \citet{swm+97}, (31)
\citet{bwpw98}, (32) \citet{dbam98}, (33) \citet{shk+98}, (34)
\citet{swm+98}, (35) \citet{tml+98}, (36) \citet{wbp+98}, (37)
\citet{rkp99}, (38) \citet{bvl00}, (39) \citet{fc00}, (40)
\citet{ffp+00}, (41) \citet{bgp+02}.}
\end{minipage}
\end{table*}
and in Fig. \ref{mike-sizes} we show the radio source size/absorption
redshift distribution compared with various absorber sizes, D$_{\rm
c}$. Although we are uncertain which of these apply to a ``typical''
DLA (e.g. \citealt{bwl+89,cc99,ck02}), the bimodal distribution of the
continuum source sizes (particularly for the detections) itself may be
indicative of a strong covering factor influence: Using the absorber
host optical identifications for this sample we see a clear
correlation between galaxy type and radio source size, where the
spirals trace the upper segment ($\theta_{\rm QSO} > 1''$) of the
distribution and the compact galaxies appear to be associated with the
smaller radio sources ($\theta_{\rm QSO} < 0.2''$). The fact that
absorption is detected only towards the extended continuum sources
when occulted by a large galaxy strongly suggests that the covering
factor is a key issue. Toward the smaller continuum sources we would
also expect spirals (together with compact galaxies) since the
covering factor would also be large. This is not
observed. \citetalias{kc02} do note that some of the identifications
may be uncertain but, consulting the literature
(e.g. \citealt{rnt+03}), we find similar identifications, although the
morphology of the $z_{\rm abs}=0.09$ DLA towards 0738+313 is unclear
\citep{coh01} (this has therefore been flagged as unknown in
Figs. \ref{kanekar}, \ref{mike-sizes} and \ref{Toverf}). The lower
segment of the low redshift bimodal distribution therefore consists of
3 dwarf/LSBs and 3 unknowns\footnote{Excluding the unknown at $0.1''$
\& $z_{\rm abs}=0.779$ (2351+456). The $0.1''$ continuum size applies
to the total extent, although the strong core flux originates within
$\approx0.03''$ \citep{pwx+95} placing this closer to the other
sources in this segment.}. That is, in this group there are as many
compact as there are unidentified morphologies, though the latter
could well turn out to be larger galaxies. Regarding the bimodal
distribution, the binary probability of finding 6 out of 6 spiral
galaxies towards sources of angular extent $\theta_{\rm QSO} > 0.5''$
and 3 (or more) out of 4 non-spirals at $\theta_{\rm QSO} < 0.5''$ is
just 0.5\,per cent. Thus, it would seem there is some evidence to
suggest a non-negligible covering factor effect in the current sample.

Furthermore, as mentioned above, all but one of the non-detections at
low redshift have yet to be identified optically, which excludes us
from estimating a probable coverage, as done for the majority of the
detections. However, in Fig. \ref{mike-sizes} we note that the low
redshift DLAs not detected in 21-cm tend to occult the largest
background sources. In fact 6 of these are located in the upper
segment of the bimodal distribution, compared to only one in the lower
segment\footnote{If applying the $\theta_{\rm QSO} = 0.5''$ cut-off this
becomes 7 versus 3.}. This strongly suggests that the low redshift
non-detections may be mainly due to inadequate coverage.

However, we emphasise one possible caveat to the above argument. Several of
the larger ($\theta_{\rm QSO} > 1''$) background sources have complex
morphologies, and in fact for 6 of the non-detections (0215+015, 0336--017,
0906+430, 0957+561, 1354+258 \& 1451--375) these may be core dominated
(Table \ref{ylva}). The fact that there is no information on the core flux
at the redshifted 21-cm frequency\footnote{These should, however, still
provide a reasonable measure of the lower frequency source size
(e.g. \protect\citealt{fpa92}).} nor on the extent of the absorbing gas
means that we cannot unambiguously attribute the non-detections purely to
low covering factors, although the smallest continuum source size is $8''$
at $\geq1.4$ GHz.

One way to estimate the extent of the absorber required to fully cover
the source is via the Compton limit of the emitting region, which
places a lower limit on the size of the emitting region according to
the upper limit of 10$^{12}$ K for the brightness temperature
(e.g. \citealt{rwt+03}). {Wolfe}, {Gawiser} \& {Prochaska} (2003)
apply this to 0201+113, where $\theta_{\rm QSO}<7$ mas (Table
\ref{ylva}), to obtain a minimum extent of 4.9 mas (or 35 pc at
$z_{\rm abs}$ for $H_{0}=75$~km~s$^{-1}$~Mpc$^{-1}$, $\Omega_{\rm
matter}=0.27$ and $\Omega_{\Lambda}=0.73$). Applying this to the rest
of the 21-cm sample gives 0.7--16 mas (or 1--130 pc at $z_{\rm abs}$),
all of which are small in comparison to $\theta_{\rm QSO}$. The
results of {Wolfe}, {Gawiser} \& {Prochaska} (2003) are interesting,
however, since the extent of 0201+113 set by the Compton limit is very
close to that set by the radio observations at 1.6 GHz ($<
2.5\times5.0$ mas$^2$, \citealt{hmp84,sbom90}). From studies of the
bolometric background radiation in DLAs, {Wolfe}, {Prochaska} \&
{Gawiser} (2003) find that these absorbers consist mainly of the
cold neutral medium (CNM, where $T\sim150$ K and $n\sim10$ \ccm)
rather than the warm neutral medium (WNM, $T\sim8000$ K and $n\sim0.2$
\ccm). This supports our argument for lower covering factors and hence
lower spin temperatures in DLAs, although based on the high covering
factor towards 0738+313 (Section \ref{prev}), \citet{lbs00,kgc01}
suggest a two phase gas model dominated by the WNM ($\approx75\%$) in
both absorbers. This is consistent with the WNM+CNM mix in nearby
dwarfs \citep{yvdl01,yvl+03}, where such low metallicity systems are
expected to reflect conditions in the early Universe, with the values
of [O/H]$\sim0.02$ to $0.2$ solar (e.g. \citealt{lci+99,lgh02})
spanning the range of typical DLA metallicities up to $z_{\rm
abs}\sim3.4$ \citep{pgw+03,cwmc03}. However, even with the warm gas
contribution, the mean harmonic mean spin temperatures of $\approx800$
K towards 0738+313 still lie in the ``cool DLA'', low redshift regime
(Figs. \ref{kanekar} \& \ref{Toverf}, below).

In 0201+113 {Wolfe}, {Gawiser} \& {Prochaska} (2003) find the CNM
density to be $n\approx6$ \ccm ~and, by assuming that the total \HI
~column density comprises several CNM (21-cm absorbing) clouds of
column densities $N\approx2\times10^{20}$ \scm, they obtain an
estimate of $\lapp10$ pc for each CNM cloud. While this argument has
the draw-back that the total \HI ~column density in the CNM phase must
be estimated, it leads to a range of possibilities for 0201+113, the
two extremes of which are:
\begin{enumerate}
  \item The $\lapp$10\,pc 21-cm absorbing clouds are aligned with the
  line-of-sight towards the $\approx40$ pc (at $z_{\rm abs}$)
  background continuum, so that the coverage is low but the total
  column density is high.  
\item The clouds are distributed with
  little overlap, giving high coverage but a low effective column
  density. 
 \end{enumerate} 
Having $f\sim1$ and $N_{\rm CNM}\ll N_{\rm
  HI}$ could account for the 21-cm non-detections, while permitting a
  large fraction of CNM gas ({Wolfe}, {Gawiser} \& {Prochaska}
  2003). This emphasises how equation \ref{enew} can at best only
  provide a column density weighted mean harmonic spin temperature
  from a complex of clouds of various column densities, velocity
  dispersions and covering factors.

\section{Conclusions}

Upon a review of the literature, we suggest that less-than-complete
covering factors may be at least partly responsible for the non-detection of
21-cm absorption in many DLAs.  Although it is as important as the
spin temperature in determining the column density of the absorber,
the covering factor is at best estimated from insufficient data, and,
more often than not, simply assumed to be unity. This assumption is
prevalent in the high redshift cases, where high spin temperatures are
claimed to be dominant, despite their being no DLA host
identifications at $z_{\rm abs}\gapp0.9$.

Although all detections of 21-cm absorption occur at $z_{\rm
abs}\leq2.04$, nearly half of those searched at these redshifts remain
undetected, thus giving an apparent mix of both low and high spin
temperatures for this sample. However, of these
non-detections, $\approx70$ per cent of the absorbers occult large background
radio sources ($\gapp1''$), thus increasing the likelihood of $f<1$.
Furthermore, 21-cm absorption tends to be detected:
\begin{enumerate} 
	\item Towards large background sources ($>1''$) only when the
DLA is known to be associated with a large galaxy.  
\item Towards compact sources ($<0.1''$) when the DLA is associated with
an LSB or dwarf galaxy.
\end{enumerate}
%p-08aug.tex has fuller argument
Naturally, we would also expect detections where the absorber is associated
with a large spiral occulting a compact background continuum, thereby
providing large coverage. However, this is not observed. Of the 6
detections towards small background sources, 3 of the hosts are
unidentified, thus not ruling out the possibility of an approximately equal
mix of large and compact DLA hosts.

From the observed bolometric background, {Wolfe}, {Gawiser} \&
{Prochaska} (2003) propose that the 21-cm absorbing CNM is a major
component of DLAs. They suggest that this is comprised of several
lower column density clouds which may exhibit a large total column
density but low covering factor when aligned along the line-of-sight
or, conversely, low individual cloud column densities and a large
covering factor, where overlap along the line-of-sight is
minimal. Therefore, when attributing the non-detection of 21-cm
absorption in DLAs to the effects of the covering factor (or spin
temperature), we should bear in mind that the total \HI ~column
density may not provide a reliable estimate of the column density of
the CNM along the line-of-sight to the background continuum. Further
complications arise from the fact that insufficient detail is known
about the flux contributions from the various components of the
background continuum illuminating the DLA.  The high resolution maps
are at different frequencies than that of redshifted 21-cm and, if the
components are extended, spectral indices may vary across them, making
estimates of the effective continuum size at the appropriate frequency
difficult. It is clear that high resolution observations at the
appropriate frequencies are required in order to fully address this
issue. Combining these with the absorption profiles at various
locations over the background continuum (as in \citealt{lbs00}) will
give far better estimates of the covering factor.

\begin{figure}
\vspace{6.2cm}
\includegraphics{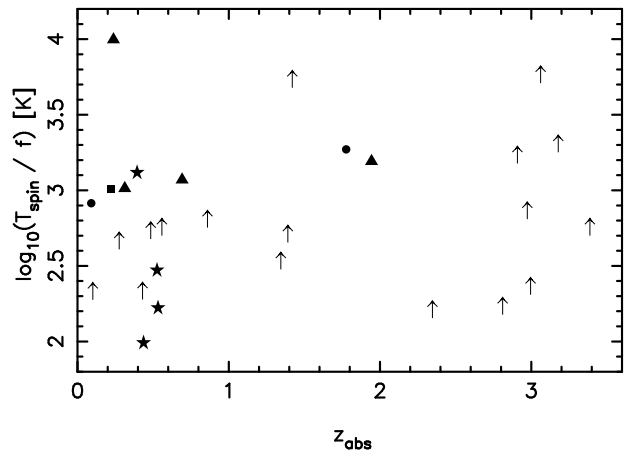}
\caption{Spin temperature/covering factor ratio against absorption
redshift for the optically thin ($\tau\lapp0.3$) 21-cm searches. Again
the FWHM of the non-detections is assumed to be 20 \kms (Section
\ref{20kms}).}
\label{Toverf}
\end{figure}

Despite the above caveats, our analysis suggests that the covering factor
may play an important r\^{o}le, thereby significantly lowering the spin
temperature estimates of the DLAs for which the covering factors have
been assumed (i.e. most of the non-detections) and perhaps it would be
more prudent to use $T_{\rm spin}/f$ versus absorption redshift
(Fig. \ref{Toverf}) rather than just $T_{\rm spin}$, for an
unknown value of $f$. The figure does in fact show that the values of
$T_{\rm spin}/f$ span similar ranges for both the detections and
non-detections with the spirals (stars) being more concentrated
towards the low $T_{\rm spin}/f$ (and $z_{\rm abs}$)
regime\footnote{Note however that the spread along the ordinate is
considerably wider than in Fig. \ref{kanekar} (top): Over 1.2 dex in
$T_{\rm spin}/f$ cf. 0.3 in $T_{\rm spin}$. Arguably this is mainly
influenced by 1229--021 at [0.39, 3.32] but also applies when we, as
\citetalias{kc02}, include the associated system 1413+135 at [0.25,
3.28].} (although, as stated previously, the majority of
host identifications are unknown). This is apparently consistent with
the interpretation of low spin temperatures in these objects
(\citetalias{kc02}), but also with high coverage by spirals, a
possibility strongly suggested by the work presented here.

The lowering of covering factor estimates would introduce more
uniformity to the spin temperatures of DLAs, while still supporting
the hypothesis that large galaxies dominate the low redshift
population \citep{bmce00} with dwarf galaxies constituting more of the
high redshift population \citep{lf03}. That is, unlike
\citetalias{kc02} who suggest that this is evident through increased
spin temperatures at high redshift, we hypothesize that the absorption
cross-sections of compact galaxies at these redshifts simply lack the
extent to effectively cover the background radio continua.

\section*{Acknowledgments}
We wish to thank the referee, Wendy Lane, whose helpful
comments improved the manuscript. Also, Matthew Whiting, Chris Blake,
Rob Beswick, Nissim Kanekar, Fredrik Rantakyr\"{o}, Sara Ellison and
Panayiotis Tzanavaris for their helpful input and comments. We also
wish to thank the John Templeton Foundation for supporting this
work. SJC gratefully acknowledges receipt of a UNSW NS Global
Fellowship and MTM is grateful to PPARC for support at the IoA under
the observational rolling grant. This research has made use of the
NASA/IPAC Extragalactic Database (NED) which is operated by the Jet
Propulsion Laboratory, California Institute of Technology, under
contract with the National Aeronautics and Space Administration.

%\bibliographystyle{mn2e}
%the above is the correct way but doesn't label, Wolfe 2003a, b
%or whatever - use apj style just now and back to mn2e when
%pasting .bbl in
%\bibliographystyle{apj} %same problem as before (m.tex)

%\bibliography{aa,ref}

\begin{thebibliography}{}

\bibitem[\protect\citeauthoryear{{Antonucci} \& {Ulvestad}}{{Antonucci} \&
  {Ulvestad}}{1985}]{au85}
{Antonucci} R.~R.~J.,  {Ulvestad} J.~S.,  1985, ApJ, 294, 158

\bibitem[\protect\citeauthoryear{{Baker}, {Mathlin}, {Churches} \&
  {Edmunds}}{{Baker} et~al.}{2000}]{bmce00}
{Baker} A.~C.,  {Mathlin} G.~P.,  {Churches} D.~K.,    {Edmunds} M.~G.,  2000,
  in Favata F.,  Kaas A.,   Wilson A.,  eds, Star Formation from the Small to
  the Large Scale, Vol.45 of ESA SP {The Chemical Evolution of the Universe}.
Noordwijk, p.~21

\bibitem[\protect\citeauthoryear{{Barthel}, {Miley}, {Schilizzi} \&
  {Lonsdale}}{{Barthel} et~al.}{1988}]{bmsl88}
{Barthel} P.~D.,  {Miley} G.~K.,  {Schilizzi} R.~T.,    {Lonsdale} C.~J.,
  1988, A\&AS, 73, 515

\bibitem[\protect\citeauthoryear{{Barthel}, {Vestergaard} \&
  {Lonsdale}}{{Barthel} et~al.}{2000}]{bvl00}
{Barthel} P.~D.,  {Vestergaard} M.,    {Lonsdale} C.~J.,  2000, A\&A, 354, 7

\bibitem[\protect\citeauthoryear{{Beasley}, {Gordon}, {Peck}, {Petrov},
  {MacMillan}, {Fomalont} \& {Ma}}{{Beasley} et~al.}{2002}]{bgp+02}
{Beasley} A.~J.,  {Gordon} D.,  {Peck} A.~B.,  {Petrov} L.,  {MacMillan} D.~S.,
   {Fomalont} E.~B.,    {Ma} C.,  2002, ApJS, 141, 13

\bibitem[\protect\citeauthoryear{{Bondi}, {Padrielli}, {Fanti}, {Ficarra},
  {Gregorini}, {Mantovani}, {Bartel}, {Romney}, {Nicolson} \& {Weiler}}{{Bondi}
  et~al.}{1996}]{bpf+96}
{Bondi} M., et al.,  1996, A\&A, 308, 415

\bibitem[\protect\citeauthoryear{{Briggs}}{{Briggs}}{1999}]{bri98}
{Briggs} F.~H.,  1999, in Carilli C.,  Radford S.,  Menton K.,   Langston G.,
  eds, Highly Redshifted Radio Lines {Redshifted 21cm Line Absorption by
  Intervening Galaxies}.
ASP Conf. Ser., p.~16

\bibitem[\protect\citeauthoryear{{Briggs}, {Brinks} \& {Wolfe}}{{Briggs}
  et~al.}{1997}]{bbw97}
{Briggs} F.~H.,  {Brinks} E.,    {Wolfe} A.~M.,  1997, AJ, 113, 467

\bibitem[\protect\citeauthoryear{Briggs \& Wolfe}{Briggs \& Wolfe}{1983}]{bw83}
Briggs F.~H.,  Wolfe A.~M.,  1983, ApJ, 268, 76

\bibitem[\protect\citeauthoryear{{Briggs}, {Wolfe}, {Liszt}, {Davis} \&
  {Turner}}{{Briggs} et~al.}{1989}]{bwl+89}
{Briggs} F.~H.,  {Wolfe} A.~M.,  {Liszt} H.~S.,  {Davis} M.~M.,    {Turner}
  K.~L.,  1989, ApJ, 341, 650

\bibitem[\protect\citeauthoryear{{Brown} \& {Mitchell}}{{Brown} \&
  {Mitchell}}{1983}]{bm83}
{Brown} R.~L.,  {Mitchell} K.~J.,  1983, ApJ, 264, 87

\bibitem[\protect\citeauthoryear{{Brown} \& {Roberts}}{{Brown} \&
  {Roberts}}{1973}]{br73}
{Brown} R.~L.,  {Roberts} M.~S.,  1973, ApJ, 184, L7

\bibitem[\protect\citeauthoryear{{Browne}, {Wilkinson}, {Patnaik} \&
  {Wrobel}}{{Browne} et~al.}{1998}]{bwpw98}
{Browne} I.~W.~A.,  {Wilkinson} P.~N.,  {Patnaik} A.~R.,    {Wrobel} J.~M.,
  1998, MNRAS, 293, 257

\bibitem[\protect\citeauthoryear{{Campbell}, {Lehar}, {Corey}, {Shapiro} \&
  {Falco}}{{Campbell} et~al.}{1995}]{clc+95}
{Campbell} R.~M.,  {Lehar} J.,  {Corey} B.~E.,  {Shapiro} I.~I.,    {Falco}
  E.~E.,  1995, AJ, 110, 2566

\bibitem[\protect\citeauthoryear{{Carilli}, {Lane}, {de Bruyn}, {Braun} \&
  {Miley}}{{Carilli} et~al.}{1996}]{cld+96}
{Carilli} C.~L.,  {Lane} W.,  {de Bruyn} A.~G.,  {Braun} R.,    {Miley} G.~K.,
  1996, AJ, 111, 1830

\bibitem[\protect\citeauthoryear{{Carilli}, {Perlman} \& {Stocke}}{{Carilli}
  et~al.}{1992}]{cps92}
{Carilli} C.~L.,  {Perlman} E.~S.,    {Stocke} J.~T.,  1992, ApJ, 400, L13

\bibitem[\protect\citeauthoryear{Chen \& Lanzetta}{Chen \&
  Lanzetta}{2003}]{cl03}
Chen H.-W.,  Lanzetta K.~M.,  2003, ApJ, 597, 706

\bibitem[\protect\citeauthoryear{{Chengalur} \& {Kanekar}}{{Chengalur} \&
  {Kanekar}}{1999}]{ck99}
{Chengalur} J.~N.,  {Kanekar} N.,  1999, MNRAS, 302, L29

\bibitem[\protect\citeauthoryear{{Chengalur} \& {Kanekar}}{{Chengalur} \&
  {Kanekar}}{2000}]{ck00}
{Chengalur} J.~N.,  {Kanekar} N.,  2000, MNRAS, 318, 303

\bibitem[\protect\citeauthoryear{{Chengalur} \& {Kanekar}}{{Chengalur} \&
  {Kanekar}}{2002}]{ck02}
{Chengalur} J.~N.,  {Kanekar} N.,  2002, A\&A, 388, 383

\bibitem[\protect\citeauthoryear{{Chu}, {B{\aa}{\aa}th}, {Rantakyr\"{o}},
  {Zhang} \& {Nicholson}}{{Chu} et~al.}{1996}]{cbr+96}
{Chu} L.~B.,  {B{\aa}{\aa}th} F.~T.,  {Rantakyr\"{o}} F.~J.,  {Zhang} H.~S.,
  {Nicholson} G.,  1996, A\&A, 307, 15

\bibitem[\protect\citeauthoryear{{Churchill} \& {Charlton}}{{Churchill} \&
  {Charlton}}{1999}]{cc99}
{Churchill} C.~W.,  {Charlton} J.~C.,  1999, AJ, 118, 59

\bibitem[\protect\citeauthoryear{{Cohen}}{{Cohen}}{2001}]{coh01}
{Cohen} J.~G.,  2001, AJ, 121, 1275

\bibitem[\protect\citeauthoryear{Curran, Murphy, Webb, Rantakyr\"{o}, Johansson
  \& Nikoli\'{c}}{Curran et~al.}{2002}]{cwn+02}
Curran S.~J.,  Murphy M.~T.,  Webb J.~K.,  Rantakyr\"{o} F.,  Johansson L.
  E.~B.,    Nikoli\'{c} S.,  2002a, A\&A, 394, 763

\bibitem[\protect\citeauthoryear{Curran, Webb, Murphy, Bandiera, Corbelli \&
  Flambaum}{Curran et~al.}{2002}]{cwbc01}
Curran S.~J.,  Webb J.~K.,  Murphy M.~T.,  Bandiera R.,  Corbelli E.,
  Flambaum V.~V.,  2002b, PASA, 19, 455

\bibitem[\protect\citeauthoryear{Curran, Webb, Murphy \& Carswell}{Curran
  et~al.}{2004}]{cwmc03}
Curran S.~J.,  Webb J.~K.,  Murphy M.~T.,    Carswell R.~F.,  2004, MNRAS, 351,
  L24

\bibitem[\protect\citeauthoryear{{Dallacasa}, {Bondi}, {Alef} \&
  {Mantovani}}{{Dallacasa} et~al.}{1998}]{dbam98}
{Dallacasa} D.,  {Bondi} M.,  {Alef} W.,    {Mantovani} F.,  1998, A\&AS, 129,
  219

\bibitem[\protect\citeauthoryear{{Darling}, {Giovanelli}, {Haynes}, {Bower} \&
  {Bolatto}}{{Darling} et~al.}{2004}]{dgh+04}
{Darling} J.,  {Giovanelli} R.,  {Haynes} M.~P.,  {Bower} G.~C.,    {Bolatto}
  A.~D.,  2004, ApJ, 613, L101

\bibitem[\protect\citeauthoryear{{de Bruyn}, {O'Dea} \& {Baum}}{{de Bruyn}
  et~al.}{1996}]{dob96}
{de Bruyn} A.~G.,  {O'Dea} C.~P.,    {Baum} S.~A.,  1996, A\&A, 305, 450

\bibitem[\protect\citeauthoryear{{Dickey} \& {Lockman}}{{Dickey} \&
  {Lockman}}{1990}]{dl90}
{Dickey} J.~M.,  {Lockman} F.~J.,  1990, ARA\&A, 28, 215

\bibitem[\protect\citeauthoryear{Ellison, Yan, Hook, Pettini, Wall \&
  Shaver}{Ellison et~al.}{2001}]{eyh+02}
Ellison S.~L.,  Yan L.,  Hook I.~M.,  Pettini M.,  Wall J.~V.,    Shaver P.,
  2001, A\&A, 379, 393

\bibitem[\protect\citeauthoryear{{Fejes}, {Porcas} \& {Akujor}}{{Fejes}
  et~al.}{1992}]{fpa92}
{Fejes} I.,  {Porcas} R.~W.,    {Akujor} C.~E.,  1992, A\&A, 257, 459

\bibitem[\protect\citeauthoryear{{Fey} \& {Charlot}}{{Fey} \&
  {Charlot}}{1997}]{fc97}
{Fey} A.~L.,  {Charlot} P.,  1997, ApJS, 111, 95

\bibitem[\protect\citeauthoryear{{Fey} \& {Charlot}}{{Fey} \&
  {Charlot}}{2000}]{fc00}
{Fey} A.~L.,  {Charlot} P.,  2000, ApJS, 128, 17

\bibitem[\protect\citeauthoryear{{Fey}, {Clegg} \& {Fomalont}}{{Fey}
  et~al.}{1996}]{fcf96}
{Fey} A.~L.,  {Clegg} A.~W.,    {Fomalont} E.~B.,  1996, ApJS, 105, 299

\bibitem[\protect\citeauthoryear{{Fomalont}, {Frey}, {Paragi}, {Gurvits},
  {Scott}, {Taylor}, {Edwards} \& {Hirabayashi}}{{Fomalont}
  et~al.}{2000}]{ffp+00}
{Fomalont} E.~B.,  {Frey} S.,  {Paragi} Z.,  {Gurvits} L.~I.,  {Scott} W.~K.,
  {Taylor} A.~R.,  {Edwards} P.~G.,    {Hirabayashi} H.,  2000, ApJS, 131, 95

\bibitem[\protect\citeauthoryear{{Gower} \& {Hutchings}}{{Gower} \&
  {Hutchings}}{1984}]{gh84}
{Gower} A.~C.,  {Hutchings} J.~B.,  1984, AJ, 89, 1658

\bibitem[\protect\citeauthoryear{{Gurvits}, {Schilizzi}, {Barthel},
  {Kardashev}, {Kellermann}, {Lobanov}, {Pauliny-Toth} \& {Popov}}{{Gurvits}
  et~al.}{1994}]{gsb+94}
{Gurvits} L.~I.,  {Schilizzi} R.~T.,  {Barthel} P.~D.,  {Kardashev} N.~S.,
  {Kellermann} K.~I.,  {Lobanov} A.~P.,  {Pauliny-Toth} I.~I.~K.,    {Popov}
  M.~V.,  1994, A\&A, 291, 737

\bibitem[\protect\citeauthoryear{Haehnelt, Steinmetz \& Rauch}{Haehnelt
  et~al.}{1998}]{hsr98}
Haehnelt M.~G.,  Steinmetz M.,    Rauch M.,  1998, ApJ, 495, 647

\bibitem[\protect\citeauthoryear{{Harvanek}, {Stocke}, {Morse} \&
  {Rhee}}{{Harvanek} et~al.}{1997}]{hsmr97}
{Harvanek} M.,  {Stocke} J.~T.,  {Morse} J.~A.,    {Rhee} G.,  1997, AJ, 114,
  2240

\bibitem[\protect\citeauthoryear{{Hintzen}, {Ulvestad} \& {Owen}}{{Hintzen}
  et~al.}{1983}]{huo83}
{Hintzen} P.,  {Ulvestad} J.,    {Owen} F.,  1983, AJ, 88, 709

\bibitem[\protect\citeauthoryear{{Hodges}, {Mutel} \& {Phillips}}{{Hodges}
  et~al.}{1984}]{hmp84}
{Hodges} M.~W.,  {Mutel} R.~L.,    {Phillips} R.~B.,  1984, AJ, 89, 1327

\bibitem[\protect\citeauthoryear{Kanekar \& Chengalur}{Kanekar \&
  Chengalur}{1997}]{kc97}
Kanekar N.,  Chengalur J.~N.,  1997, MNRAS, 292, 831

\bibitem[\protect\citeauthoryear{Kanekar \& Chengalur}{Kanekar \&
  Chengalur}{2001}]{kc01a}
Kanekar N.,  Chengalur J.~N.,  2001, A\&A, 369, 42

\bibitem[\protect\citeauthoryear{Kanekar \& Chengalur}{Kanekar \&
  Chengalur}{2003}]{kc02}
Kanekar N.,  Chengalur J.~N.,  2003, A\&A, 399, 857

\bibitem[\protect\citeauthoryear{{Kanekar}, {Chengalur}, {Subrahmanyan} \&
  {Petitjean}}{{Kanekar} et~al.}{2001}]{kcsp01}
{Kanekar} N.,  {Chengalur} J.~N.,  {Subrahmanyan} R.,    {Petitjean} P.,  2001,
  A\&A, 367, 46

\bibitem[\protect\citeauthoryear{Kanekar, Ghosh \& Chengalur}{Kanekar
  et~al.}{2001}]{kgc01}
Kanekar N.,  Ghosh T.,    Chengalur J.~N.,  2001, A\&A, 373, 394

\bibitem[\protect\citeauthoryear{Lane, Smette, Briggs, Rao, Turnshek \&
  Meylan}{Lane et~al.}{1998}]{lsb+98}
Lane W.,  Smette A.,  Briggs F.~H.,  Rao S.~M.,  Turnshek D.~A.,    Meylan G.,
  1998, AJ, 116, 26

\bibitem[\protect\citeauthoryear{Lane \& Briggs}{Lane \& Briggs}{2001}]{lb01}
Lane W.~M.,  Briggs F.~H.,  2001, ApJ, 561, L27

\bibitem[\protect\citeauthoryear{{Lane}, {Briggs} \& {Smette}}{{Lane}
  et~al.}{2000}]{lbs00}
{Lane} W.~M.,  {Briggs} F.~H.,    {Smette} A.,  2000, ApJ, 532, 146

\bibitem[\protect\citeauthoryear{{Lanfranchi} \& {Fria{\c c}a}}{{Lanfranchi} \&
  {Fria{\c c}a}}{2003}]{lf03}
{Lanfranchi} G.~A.,  {Fria{\c c}a} A.~C.~S.,  2003, MNRAS, 343, 481

\bibitem[\protect\citeauthoryear{{Lanzetta}, {Wolfe} \& {Turnshek}}{{Lanzetta}
  et~al.}{1995}]{lwt95}
{Lanzetta} K.~M.,  {Wolfe} A.~M.,    {Turnshek} D.~A.,  1995, ApJ, 440, 435

\bibitem[\protect\citeauthoryear{{Lanzetta}, {Wolfe}, {Turnshek}, {Lu},
  {McMahon} \& {Hazard}}{{Lanzetta} et~al.}{1991}]{lwt+91}
{Lanzetta} K.~M.,  {Wolfe} A.~M.,  {Turnshek} D.~A.,  {Lu} L.,  {McMahon}
  R.~G.,    {Hazard} C.,  1991, ApJS, 77, 1

\bibitem[\protect\citeauthoryear{{Le Brun}, Bergeron, Boiss\'{e} \&
  Deharveng}{{Le Brun} et~al.}{1997}]{lbbd97}
{Le Brun} V.,  Bergeron J.,  Boiss\'{e} P.,    Deharveng J.~M.,  1997, A\&A,
  321, 733

\bibitem[\protect\citeauthoryear{{Lee}, {Grebel} \& {Hodge}}{{Lee}
  et~al.}{2002}]{lgh02}
{Lee} H.,  {Grebel} E.~K.,    {Hodge} P.~W.,  2002, BAAS, 34, 1121

\bibitem[\protect\citeauthoryear{{Levshakov} \& {Varshalovich}}{{Levshakov} \&
  {Varshalovich}}{1985}]{lv85}
{Levshakov} S.~A.,  {Varshalovich} D.~A.,  1985, MNRAS, 212, 517

\bibitem[\protect\citeauthoryear{{Lipovetsky}, {Chaffee}, {Izotov}, {Foltz},
  {Kniazev} \& {Hopp}}{{Lipovetsky} et~al.}{1999}]{lci+99}
{Lipovetsky} V.~A.,  {Chaffee} F.~H.,  {Izotov} Y.~I.,  {Foltz} C.~B.,
  {Kniazev} A.~Y.,    {Hopp} U.,  1999, ApJ, 519, 177

\bibitem[\protect\citeauthoryear{{Lister}, {Gower} \& {Hutchings}}{{Lister}
  et~al.}{1994}]{lgh94}
{Lister} M.~L.,  {Gower} A.~C.,    {Hutchings} J.~B.,  1994, AJ, 108, 821

\bibitem[\protect\citeauthoryear{{Lonsdale}, {Barthel} \& {Miley}}{{Lonsdale}
  et~al.}{1993}]{lbm93}
{Lonsdale} C.~J.,  {Barthel} P.~D.,    {Miley} G.~K.,  1993, ApJS, 87, 63

\bibitem[\protect\citeauthoryear{{Murphy}, {Browne} \& {Perley}}{{Murphy}
  et~al.}{1993}]{mbp93}
{Murphy} D.~W.,  {Browne} I.~W.~A.,    {Perley} R.~A.,  1993, MNRAS, 264, 298

\bibitem[\protect\citeauthoryear{{Neff} \& {Hutchings}}{{Neff} \&
  {Hutchings}}{1990}]{nh90}
{Neff} S.~G.,  {Hutchings} J.~B.,  1990, AJ, 100, 1441

\bibitem[\protect\citeauthoryear{{Perley}}{{Perley}}{1982}]{per82}
{Perley} R.~A.,  1982, AJ, 87, 859

\bibitem[\protect\citeauthoryear{{Perlman}, {Stocke}, {Shaffer}, {Carilli} \&
  {Ma}}{{Perlman} et~al.}{1994}]{pss+94}
{Perlman} E.~S.,  {Stocke} J.~T.,  {Shaffer} D.~B.,  {Carilli} C.~L.,    {Ma}
  C.,  1994, ApJ, 424, L69

\bibitem[\protect\citeauthoryear{{Pihlstr{\" o}m}, {Conway} \&
  {Vermeulen}}{{Pihlstr{\" o}m} et~al.}{2003}]{pcv03}
{Pihlstr{\" o}m} Y.~M.,  {Conway} J.~E.,    {Vermeulen} R.~C.,  2003, A\&A,
  404, 871

\bibitem[\protect\citeauthoryear{{Pihlstr{\" o}m}, {Vermeulen}, {Taylor} \&
  {Conway}}{{Pihlstr{\" o}m} et~al.}{1999}]{pvtc99}
{Pihlstr{\" o}m} Y.~M.,  {Vermeulen} R.~C.,  {Taylor} G.~B.,    {Conway} J.~E.,
   1999, ApJ, 525, L13

\bibitem[\protect\citeauthoryear{{Polatidis}, {Wilkinson}, {Xu}, {Readhead},
  {Pearson}, {Taylor} \& {Vermeulen}}{{Polatidis} et~al.}{1995}]{pwx+95}
{Polatidis} A.~G.,  {Wilkinson} P.~N.,  {Xu} W.,  {Readhead} A.~C.~S.,
  {Pearson} T.~J.,  {Taylor} G.~B.,    {Vermeulen} R.~C.,  1995, ApJS, 98, 1

\bibitem[\protect\citeauthoryear{{Price}, {Gower}, {Hutchings}, {Talon},
  {Duncan} \& {Ross}}{{Price} et~al.}{1993}]{pgh+93}
{Price} R.,  {Gower} A.~C.,  {Hutchings} J.~B.,  {Talon} S.,  {Duncan} D.,
  {Ross} G.,  1993, ApJS, 86, 365

\bibitem[\protect\citeauthoryear{Prochaska, Gawiser, Wolfe, Castro \&
  Djorgovski}{Prochaska et~al.}{2003}]{pgw+03}
Prochaska J.~X.,  Gawiser E.,  Wolfe A.~M.,  Castro S.,    Djorgovski S.~G.,
  2003, ApJ, 595, L9

\bibitem[\protect\citeauthoryear{{Prochaska} \& {Wolfe}}{{Prochaska} \&
  {Wolfe}}{1997}]{pw97}
{Prochaska} J.~X.,  {Wolfe} A.~M.,  1997, ApJ, 487, 73

\bibitem[\protect\citeauthoryear{{Rantakyr{\" o}}, {Wiik}, {Tornikoski},
  {Valtaoja} \& {B{\aa}{\aa}th}}{{Rantakyr{\" o}} et~al.}{2003}]{rwt+03}
{Rantakyr{\" o}} F.~T.,  {Wiik} K.,  {Tornikoski} M.,  {Valtaoja} E.,
  {B{\aa}{\aa}th} L.~B.,  2003, A\&A, 405, 473

\bibitem[\protect\citeauthoryear{Rao, Nestor, Turnshek, Lane, Monier \&
  Bergeron}{Rao et~al.}{2003}]{rnt+03}
Rao S.,  Nestor D.~B.,  Turnshek D.,  Lane W.~M.,  Monier E.~M.,    Bergeron
  J.,  2003, ApJ, 595, 94

\bibitem[\protect\citeauthoryear{{Rao} \& {Turnshek}}{{Rao} \&
  {Turnshek}}{1998}]{rt98}
{Rao} S.~M.,  {Turnshek} D.~A.,  1998, ApJ, 500, L115

\bibitem[\protect\citeauthoryear{{Reid}, {Shone}, {Akujor}, {Browne}, {Murphy},
  {Pedelty}, {Rudnick} \& {Walsh}}{{Reid} et~al.}{1995}]{rsa+95}
{Reid} A.,  {Shone} D.~L.,  {Akujor} C.~E.,  {Browne} I.~W.~A.,  {Murphy}
  D.~W.,  {Pedelty} J.,  {Rudnick} L.,    {Walsh} D.,  1995, A\&AS, 110, 213

\bibitem[\protect\citeauthoryear{{Reid}, {Kronberg} \& {Perley}}{{Reid}
  et~al.}{1999}]{rkp99}
{Reid} R.~I.,  {Kronberg} P.~P.,    {Perley} R.~A.,  1999, ApJS, 124, 285

\bibitem[\protect\citeauthoryear{{Roberts}, {Brown}, {Brundage}, {Rots},
  {Haynes} \& {Wolfe}}{{Roberts} et~al.}{1976}]{rbb+76}
{Roberts} M.~S.,  {Brown} R.~L.,  {Brundage} W.~D.,  {Rots} A.~H.,  {Haynes}
  M.~P.,    {Wolfe} A.~M.,  1976, AJ, 81, 293

\bibitem[\protect\citeauthoryear{{Rogora}, {Padrielli} \& {de Ruiter}}{{Rogora}
  et~al.}{1987}]{rpd87}
{Rogora} A.,  {Padrielli} L.,    {de Ruiter} H.~R.,  1987, A\&AS, 67, 267

\bibitem[\protect\citeauthoryear{Ryan-Weber, Webster \&
  Staveley-Smith}{Ryan-Weber et~al.}{2003}]{rws03}
Ryan-Weber E.~V.,  Webster R.~L.,    Staveley-Smith L.,  2003, MNRAS, 343, 1195

\bibitem[\protect\citeauthoryear{{Saikia}, {Holmes}, {Kulkarni}, {Salter} \&
  {Garrington}}{{Saikia} et~al.}{1998}]{shk+98}
{Saikia} D.~J.,  {Holmes} G.~F.,  {Kulkarni} A.~R.,  {Salter} C.~J.,
  {Garrington} S.~T.,  1998, MNRAS, 298, 877

\bibitem[\protect\citeauthoryear{{Schilizzi}, {Kapahi} \& {Neff}}{{Schilizzi}
  et~al.}{1982}]{skn82}
{Schilizzi} R.~T.,  {Kapahi} V.~K.,    {Neff} S.~G.,  1982, JA\&A, 3, 173

\bibitem[\protect\citeauthoryear{{Shen}, {Wan}, {Moran}, {Jauncey}, {Reynolds}
  \& {Tzioumis}}{{Shen} et~al.}{1997}]{swm+97}
{Shen} Z.-Q.,  {Wan} T.-S.,  {Moran} J.~M.,  {Jauncey} D.~L.,  {Reynolds}
  J.~E.,    {Tzioumis} A.~K.,  1997, AJ, 114, 1999

\bibitem[\protect\citeauthoryear{{Shen}, {Wan}, {Moran}, {Jauncey}, {Reynolds}
  \& {Tzioumis}}{{Shen} et~al.}{1998}]{swm+98}
{Shen} Z.-Q.,  {Wan} T.-S.,  {Moran} J.~M.,  {Jauncey} D.~L.,  {Reynolds}
  J.~E.,    {Tzioumis} A.~K.,  1998, AJ, 115, 1357

\bibitem[\protect\citeauthoryear{{Srianand} \& {Petitjean}}{{Srianand} \&
  {Petitjean}}{1998}]{sp98}
{Srianand} R.,  {Petitjean} P.,  1998, A\&A, 335, 33

\bibitem[\protect\citeauthoryear{{Stanghellini}, {Baum}, {O'Dea} \&
  {Morris}}{{Stanghellini} et~al.}{1990}]{sbom90}
{Stanghellini} C.,  {Baum} S.~A.,  {O'Dea} C.~P.,    {Morris} G.~B.,  1990,
  A\&A, 233, 379

\bibitem[\protect\citeauthoryear{{Stanghellini}, {O'Dea}, {Baum}, {Dallacasa},
  {Fanti} \& {Fanti}}{{Stanghellini} et~al.}{1997}]{sob+97}
{Stanghellini} C.,  {O'Dea} C.~P.,  {Baum} S.~A.,  {Dallacasa} D.,  {Fanti} R.,
     {Fanti} C.,  1997, A\&A, 325, 943

\bibitem[\protect\citeauthoryear{Stanimirovic, Weisberg, Dickey, de~la Fuente,
  Devine, Hedden \& Anderson}{Stanimirovic et~al.}{2003}]{swd+03a}
Stanimirovic S.,  Weisberg J.~M.,  Dickey J.~M.,  de~la Fuente A.,  Devine K.,
  Hedden A.,    Anderson S.~B.,  2003, in Magnetic Fields and Star Formation:
  Theory Versus Observations PSR B1849+00 probes the tiny-scale molecular gas?
Kluwer

\bibitem[\protect\citeauthoryear{{Stocke}, {Foltz}, {Weymann} \&
  {Christiansen}}{{Stocke} et~al.}{1984}]{sfwc84}
{Stocke} J.~T.,  {Foltz} C.~B.,  {Weymann} R.~J.,    {Christiansen} W.~A.,
  1984, ApJ, 280, 476

\bibitem[\protect\citeauthoryear{{Tingay}, {Murphy}, {Lovell}, {Costa},
  {McCulloch}, {Edwards}, {Jauncey}, {Reynolds}, {Tzioumis}, {King}, {Jones},
  {Preston}, {Meier}, {van Ommen}, {Nicolson} \& {Quick}}{{Tingay}
  et~al.}{1998}]{tml+98}
{Tingay} S.~J.,  et al,  1998, ApJ, 497, 594

\bibitem[\protect\citeauthoryear{{Turnshek}, {Wolfe}, {Lanzetta}, {Briggs},
  {Cohen}, {Foltz}, {Smith} \& {Wilkes}}{{Turnshek} et~al.}{1989}]{twl+89}
{Turnshek} D.~A.,  {Wolfe} A.~M.,  {Lanzetta} K.~M.,  {Briggs} F.~H.,  {Cohen}
  R.~D.,  {Foltz} C.~B.,  {Smith} H.~E.,    {Wilkes} B.~J.,  1989, ApJ, 344,
  567

\bibitem[\protect\citeauthoryear{{Ulvestad}, {Johnston}, {Perley} \&
  {Fomalont}}{{Ulvestad} et~al.}{1981}]{ujpf81}
{Ulvestad} J.,  {Johnston} K.,  {Perley} R.,    {Fomalont} E.,  1981, AJ, 86,
  1010

\bibitem[\protect\citeauthoryear{{van Breugel}, {Fanti}, {Fanti},
  {Stanghellini}, {Schilizzi} \& {Spencer}}{{van Breugel}
  et~al.}{1992}]{vff+92}
{van Breugel} W.~J.~M.,  {Fanti} C.,  {Fanti} R.,  {Stanghellini} C.,
  {Schilizzi} R.~T.,    {Spencer} R.~E.,  1992, A\&A, 256, 56

\bibitem[\protect\citeauthoryear{Vermeulen, Pihlstr\"{o}m, Tschager, de Vries,
  Conway, Barthel, Baum, Braun, Bremer, Miley, O'Dea, Roettgering, Schilizzi,
  Snellen \& Taylor}{Vermeulen et~al.}{2003}]{vpt+03}
Vermeulen R.~C.,  et al.,  2003, A\&A, 404, 861

\bibitem[\protect\citeauthoryear{{Wilkinson}, {Browne}, {Patnaik}, {Wrobel} \&
  {Sorathia}}{{Wilkinson} et~al.}{1998}]{wbp+98}
{Wilkinson} P.~N.,  {Browne} I.~W.~A.,  {Patnaik} A.~R.,  {Wrobel} J.~M.,
  {Sorathia} B.,  1998, MNRAS, 300, 790

\bibitem[\protect\citeauthoryear{{Wills}, {Thompson}, {Han}, {Netzer}, {Wills},
  {Baldwin}, {Ferland}, {Browne} \& {Brotherton}}{{Wills}
  et~al.}{1995}]{wth+95}
{Wills} B.~J., et al., 1995, ApJ, 447, 139

\bibitem[\protect\citeauthoryear{{Wolfe}}{{Wolfe}}{1980}]{wol80}
{Wolfe} A.~M.,  1980, Physica Scripta, 21, 744

\bibitem[\protect\citeauthoryear{Wolfe, Briggs \& Jauncey}{Wolfe
  et~al.}{1981}]{wbj81}
Wolfe A.~M.,  Briggs F.~H.,    Jauncey D.~L.,  1981, ApJ, 248, 460

\bibitem[\protect\citeauthoryear{{Wolfe}, {Briggs}, {Turnshek}, {Davis},
  {Smith} \& {Cohen}}{{Wolfe} et~al.}{1985}]{wbt+85}
{Wolfe} A.~M.,  {Briggs} F.~H.,  {Turnshek} D.~A.,  {Davis} M.~M.,  {Smith}
  H.~E.,    {Cohen} R.~D.,  1985, ApJ, 294, L67

\bibitem[\protect\citeauthoryear{{Wolfe} \& {Burbidge}}{{Wolfe} \&
  {Burbidge}}{1975}]{wb75}
{Wolfe} A.~M.,  {Burbidge} G.~R.,  1975, ApJ, 200, 548

\bibitem[\protect\citeauthoryear{Wolfe \& Davis}{Wolfe \& Davis}{1979}]{wd79}
Wolfe A.~M.,  Davis M.~M.,  1979, AJ, 84, 699

\bibitem[\protect\citeauthoryear{{Wolfe}, {Gawiser} \& {Prochaska}}{{Wolfe}
  et~al.}{2003}]{wgp03}
{Wolfe} A.~M.,  {Gawiser} E.,    {Prochaska} J.~X.,  2003, ApJ, 593, 235

\bibitem[\protect\citeauthoryear{{Wolfe}, {Prochaska} \& {Gawiser}}{{Wolfe}
  et~al.}{2003}]{wpg03}
{Wolfe} A.~M.,  {Prochaska} J.~X.,    {Gawiser} E.,  2003, ApJ, 593, 215

\bibitem[\protect\citeauthoryear{{Young}, {van Zee}, {Dohm-Palmer} \&
  {Lo}}{{Young} et~al.}{2001}]{yvdl01}
{Young} L.~M.,  {van Zee} L.,  {Dohm-Palmer} R.~C.,    {Lo} K.~Y.,  2001, in
  Hibbard J.~E.,  R. M.,   van Gorkom J.~H.,  eds, Gas and Galaxy Evolution
  {Star Formation and the Interstellar Medium in Dwarf Galaxies}.
ASP Conf. Ser., Vol 240, San Francisco, p.~187

\bibitem[\protect\citeauthoryear{{Young}, {van Zee}, {Lo}, {Dohm-Palmer} \&
  {Beierle}}{{Young} et~al.}{2003}]{yvl+03}
{Young} L.~M.,  {van Zee} L.,  {Lo} K.~Y.,  {Dohm-Palmer} R.~C.,    {Beierle}
  M.~E.,  2003, ApJ, 592, 111

\end{thebibliography}
%\expandafter\ifx\csname natexlab\endcsname\relax\def\natexlab#1{#1}\fi
%need for mn2e to work properly

\bsp_small

\section*{APPENDIX A}

\label{ssec:obs}
%\subsection*{Observations and data reduction}\label{ssec:obs}

Our initial motivation for this work was a search for 21-cm
absorption in DLAs with the Parkes radio telescope.
From the September 2001 version of our catalogue of all known DLAs
\citep{cwbc01}\footnote{A version of this catalogue is continually
updated on-line and is available from
http://www.phys.unsw.edu.au/$\sim$sjc/dla} we shortlisted those which
are illuminated by radio-loud QSOs (i.e. those with a measured radio
flux density $\gapp0.1$ Jy). From these, the DLAs with redshifts
appropriate to the Parkes 70-cm receiver bandwidth (i.e. $2.1 \lapp z
\lapp 2.4$) known to occult southern ($\delta<30$\dg) QSOs were
selected (Table \ref{t1}).
\begin{table}
\centering
%\begin{minipage}{84mm}
\caption{The known southern radio illuminated DLAs and sub-DLAs which
fall into the Parkes 70-cm receiver band (B1950.0 names are used
throughout this paper). These are obtained from
\protect\citet{twl+89}$^{a}$, \protect\citet{lwt+91}$^{b}$ and
\protect\citet{eyh+02}$^{c}$, where $N_{\rm HI}$ and $z_{\rm abs}$ are
the column density (\scm) and the redshift of the DLA,
respectively. $\nu_{\rm obs}$ is the observed frequency of the
redshifted 21-cm line (MHz) and $S$ is the flux density of the
background QSO at this frequency in Jy (see
\protect\citealt{cwbc01}). For 0528--2505 the value is the flux at 372
MHz \citep{cld+96} and for 1228--113 the value is from a 327~MHz Giant
Metrewave Radio Telescope (GMRT) observation (Nissim Kanekar, private
communication). \label{t1}}
%by one of the authors (NK). \label{t1}}
\begin{tabular}{@{}l c cl c c @{}} 
\hline
QSO & $\log N_{\rm HI}$  & $z_{\rm abs}$& $\nu_{\rm obs}$& $S$ \\
\hline
0432--440$^{c}$ & 20.8 & 2.297 &  430.82 & 0.72 \\ 
0438--436$^{c}$ &  20.8 & 2.347 & 424.38 & 8.10 \\
0528--250$^{b}$  & 20.6 & 2.1410&  452.65& 0.14 \\
1017+109$^{b}$  & 19.9 & 2.380 & 420.24 & 1.43\\ 
1021--006$^{a,b}$  &  19.6 & 2.398 & 418.01 & 0.50 \\ 
1228--113$^{c}$  &  20.6 & 2.193 &444.85 & 0.35 \\
2136+141$^{a}$ & 19.8 & 2.134 & 453.22 & 0.76\\ 
\hline
\end{tabular}

%\end{minipage}
\end{table}
Although there are 7 possible targets, we used the $90$ hours
scheduled to concentrate on 3 of the 4 confirmed DLAs; 0432--440,
0438--436 \& 1228--113\footnote{1017+109, 1021--006 and 2136+141 are
candidate DLAs which have yet to be confirmed using high resolution
optical spectroscopy.}. 0528--250 was excluded due to its relatively low flux
density. The redshifts of these DLAs are
known to within $\Delta z\approx\pm0.002$
(Sara Ellison, private communication), corresponding to 
$\Delta\nu\approx\pm0.25$ MHz ($\approx\pm200$ \kms).

The observations were undertaken in January 2002 using the 70-cm
receiver on the Parkes 64-m antenna. We used the AT conversion system
with the multibeam correlator, giving a bandwidth of 8 MHz ($\Delta
z\approx\pm0.03$) over 2048 channels (i.e. a spectral dispersion of
2.8 \kms ~per channel) and position switching with 1\dg ~beam
throw. The weather was clear and dry giving system temperatures
typically between 60 and 150\,K. Although we restricted our
observations to night time, we experienced severe radio-frequency
interference (RFI) in this band. The RFI consisted primarily of sharp
(${\rm FWHM}\!\la\!6{\rm \,km\,s}$) `spikes' and broad (${\rm
FWHM}\!\ga\!100{\rm \,km\,s}^{-1}$) `humps'. The humps in particular
were found to shift in frequency between 120-s scans of the on- and
off-source positions. Therefore, attempts to subtract the off-source
scans from their corresponding on-source scans merely introduced
strong RFI variations in the residual spectrum with characteristic
length scales $\approx\!30{\rm \,km\,s}^{-1}$, i.e.~the very scales
typical of H{\sc i} 21-cm absorption lines in DLAs (see Table 2). The
stronger spikes were also often accompanied by a powerful `ringing'
such that strong correlations between neighbouring channels could be
seen to die away over $\approx\!300{\rm \,km\,s}^{-1}$ scales. The
data reduction was therefore carried out using software written
specifically for the purpose of dealing with this RFI. The reduction
steps are summarised as follows:
\begin{enumerate}
  \item The off-source scans were discarded and the
on-source scans were low-pass filtered to provide a basic continuum shape
which included the humps but ignored the spikes. The filtering scale
was set so that $\approx\!50{\rm \,km\,s}^{-1}$ absorption features could be
recovered reliably. This continuum was then subtracted from the on-source
scans leaving relatively flat residual spectra contaminated by the
spikes.
  \item For each QSO, all $\ga\!500$ contributing scans were corrected
to the heliocentric frame and re-dispersed to a common frequency scale with
a dispersion of $10{\rm \,km\,s}^{-1}$.
  \item A $1\,\sigma$ error array was constructed for each
re-dispersed scan based on the rms in a sliding 20-channel window and
channels with fluxes deviating by more than $3\sigma$ were flagged and
ignored in further reduction steps.
 \item  Finally, the
fluxes in corresponding channels in each scan were median filtered, values
more than $2\,\sigma$ from the median rejected and the remaining values
combined to form the weighted mean flux density and $1\,\sigma$ error
arrays.
\end{enumerate}

%\subsection{Data reduction}\label{subsec:red}

\begin{figure}
\includegraphics[width=\columnwidth]{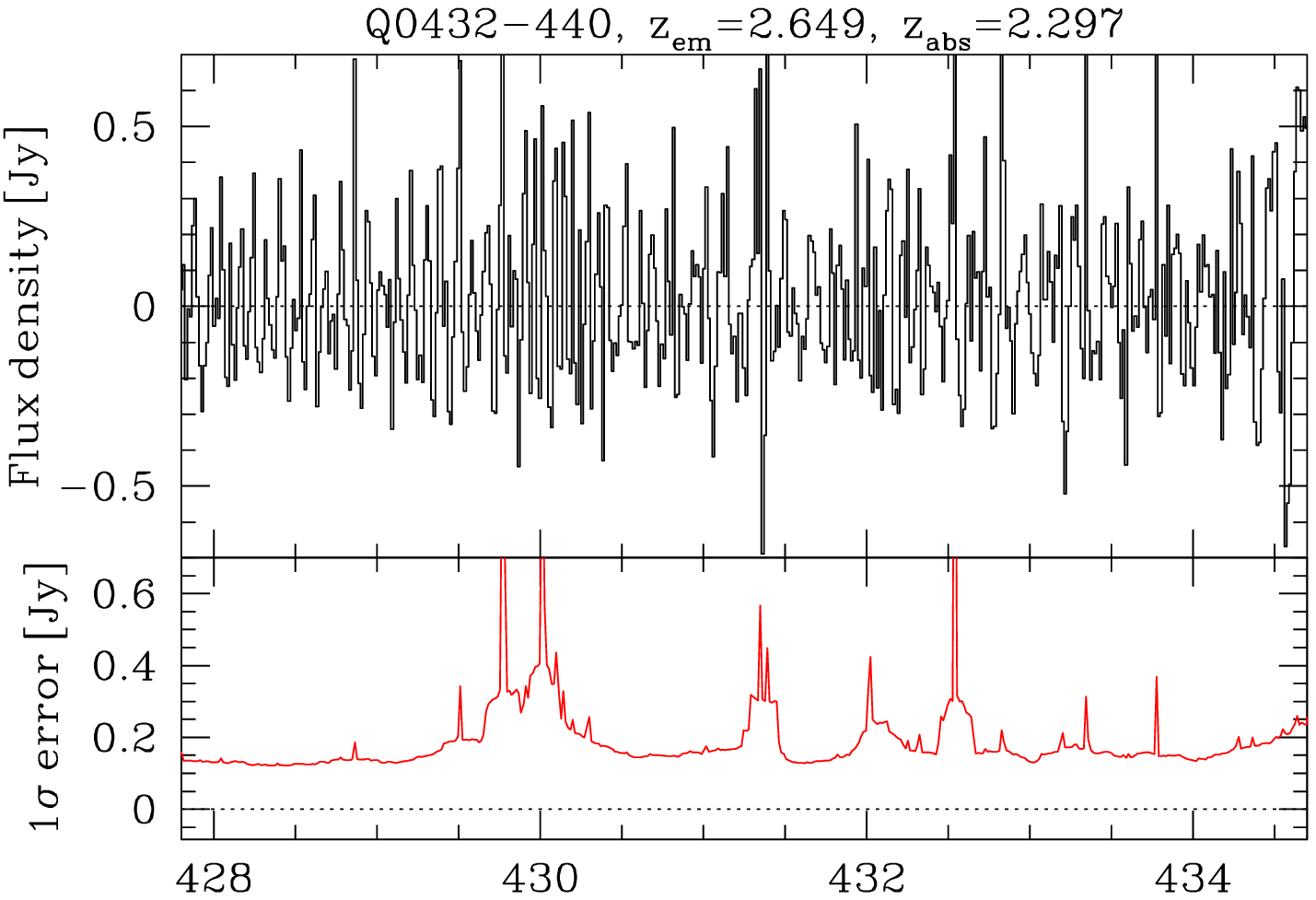}
%\vspace*{5mm}
\includegraphics[width=\columnwidth]{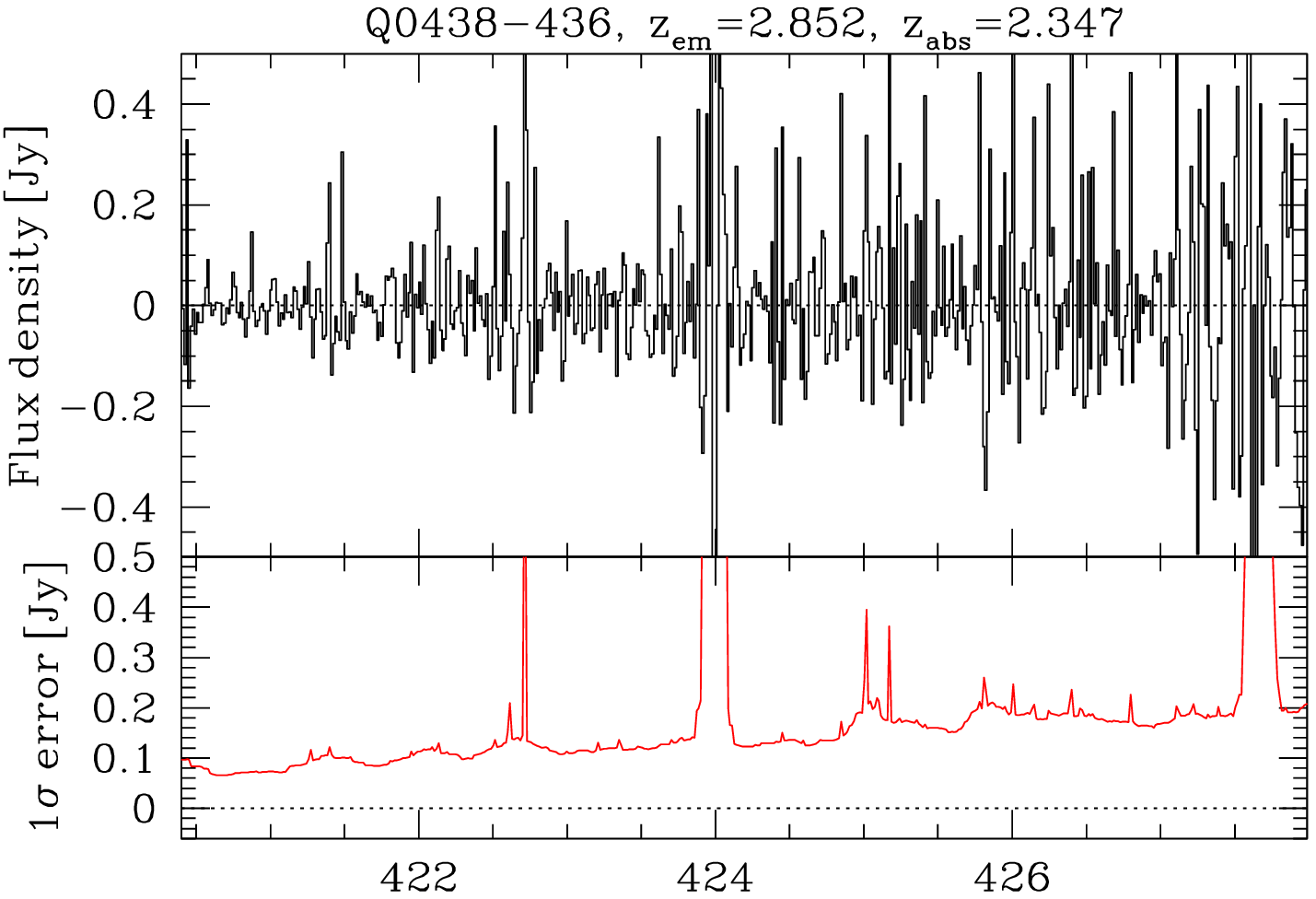}
%\vspace*{5mm}
\includegraphics[width=\columnwidth]{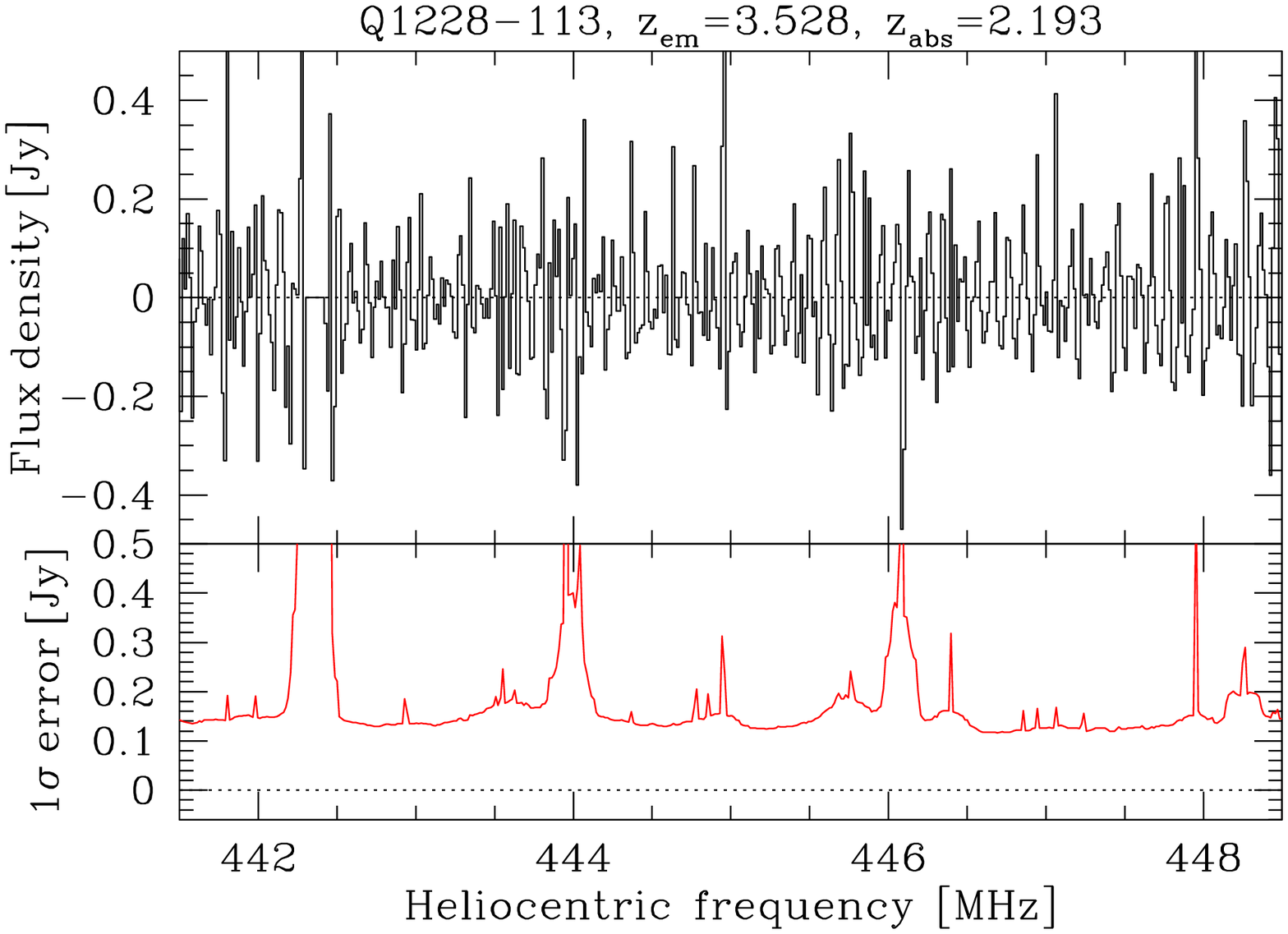}
	\caption{Parkes 21-cm spectra towards 0432--440 (top),
	0438--436 (middle) and 1228--113 (bottom). An uncertainty of
	$\Delta z\approx\pm0.002$ corresponds to the ranges
	430.6--431.1, 424.1--424.6 and 444.6--445.1 MHz, respectively,
	indicating that spikes have been successfully removed over
	the ranges expected from the optical absorption.}
\label{f1}
\end{figure}

In Fig. \ref{f1} we show the resulting spectra for each DLA. It is
clear that there are no significant, convincing absorption features
over the mitigated intervals covered by the spectra (which include
the frequencies expected from the optical redshifts). Generally, the
noise level is quite high, $\approx\!0.1$--$0.2{\rm \,Jy}$ (cf. the
theoretical value of $\sigma\lapp3$ mJy per 10 \kms ~channel) and some
regions of the spectra were so severely affected by RFI that the
errors are higher still. Where no reliable data could be gathered from
any of the contributing scans, we have assigned a normalized flux
density of zero and an effectively infinite error. Note also that the
low-pass filter applied in stage (i) of the reduction procedure
effectively prevents us from detecting absorption features with ${\rm
FWHM}\!\ga\!100{\rm \,km\,s}$. The $3\,\sigma$ optical depth limits
quoted in Table~\ref{t2} for these sources must therefore carry these
caveats.

Due to the severe RFI, no flux densities could be obtained for our
sources. Using the values in Table \ref{t1}, for $\sigma=0.2$ Jy at 10 \kms
~resolution we obtain no $3\sigma$ detection of the flux for either
0432--440 nor 1228--113 and $\tau<0.1$ for the optical depth of 21-cm
absorption in the DLA towards 0438--436. From equation \ref{e1} and the
column density of the Lyman-$\alpha$ line (Table \ref{t2}), this gives
$\frac{T_{\rm spin}}{f}\times{\rm FWHM}\gapp3500$ K \kms,
i.e. $\frac{T_{\rm spin}}{f}\gapp1200$ K per 3 \kms ~channel or $T_{\rm
s}\gapp200$ K for a FWHM of 20 \kms~(see Section \ref{20kms}).

\label{lastpage}

\end{document}